\documentclass[12pt]{article}

\usepackage{amstext,amsmath,amssymb,amsfonts,bbm,slashed}
\usepackage[latin1]{inputenc}
\usepackage{epsfig}
\usepackage{hyperref}
\usepackage{amsthm}
\usepackage{subfigure}
\usepackage{color}
\usepackage{multirow}

\usepackage{cancel}

\newcommand{\kin}{\text{kin\,}}
\newcommand{\ext}{\text{ext\,}}
\newcommand{\inter}{\text{int\,}}

\newcommand{\ren}{\text{ren\,}}

\def\N{{\mathbbm N}}
\def\Z{{\mathbbm Z}}

\newcommand{\cG}{{\mathcal G}}
\newcommand{\bG}{{\partial\mathcal G}}
\newcommand{\tJ}{{\widetilde{J}}}

\newcommand{\cexG}{\mathcal G_{\text{color}}}
\newcommand{\bJ}{ J_{\partial} }

\newcommand{\cL}{{\mathcal L}}
\newcommand{\cV}{{\mathcal V}}
\newcommand{\cF}{{\mathcal F}}

\newcommand{\cA}{{\mathcal A}}

\newtheorem{lemma}{Lemma}

\newtheorem{definition}{Definition}
\newtheorem{theorem}{Theorem}
\newtheorem{proposition}{Proposition}

\newcommand{\bea}{\begin{eqnarray}}
\newcommand{\eea}{\end{eqnarray}}
\newcommand{\beq}{\begin{equation}}
\newcommand{\eeq}{\end{equation}}

\makeatletter

\begin{document}

\begin{titlepage}
\begin{flushright}
pi-qg-252\\
ICMPA-MPA/2011/018
\end{flushright}

\vspace{20pt}

\begin{center}

{\Large\bf 3D Tensor Field Theory: \\
\medskip
Renormalization and One-loop $\beta$-functions
}
\vspace{15pt}

{\large Joseph Ben Geloun$^{a,b,\dag}$ and Dine Ousmane Samary $^{b,\ddag} $}

\vspace{15pt}

$^{a}${\sl Perimeter Institute for Theoretical Physics}\\
{\sl 31 Caroline St. N., ON, N2L 2Y5, Waterloo, Canada}\\
\vspace{5pt}

$^{b}${\sl International Chair in Mathematical Physics and Applications\\ (ICMPA-UNESCO Chair), University of Abomey-Calavi,\\
072B.P.50, Cotonou, Rep. of Benin}\\
\vspace{5pt}
E-mails:  {\sl $^{\dag}$jbengeloun@perimeterinstitute.ca, 
$^\ddag$ousmanesamarydine@yahoo.fr}

\vspace{10pt}

\begin{abstract}
We prove that the rank 3 analogue 
of the tensor model defined in 
[arXiv:1111.4997 [hep-th]] is renormalizable at all orders
of perturbation. The proof is given in the momentum 
space. The one-loop  $\gamma$- and $\beta$-functions
of the model are also determined. We find that the model
with a unique coupling constant for all interactions 
and a unique wave function
renormalization is asymptotically free in the UV. 
\end{abstract}

\end{center}

\noindent  Pacs numbers:  11.10.Gh, 04.60.-m
\\

\setcounter{footnote}{0}

\end{titlepage}

\section{Introduction}

Approaches to one of the most important problems
in physics, namely the quantum gravity (QG) conundrum, 
have evolved a lot since the last two decades. 
The most known contender having some undeniable results 
is certainly String Theory  \cite{Zwiebach:2004tj} whereas alternative approaches\footnote{We include Supergravity and M-Theory
within the String approach as theories invoking extra symmetries or dimensions.} 
like Asymptotic Safety scenario \cite{Niedermaier:2006wt}, 
Noncommutative Geometry \cite{Connes:1994yd}, 
 Dynamical Triangulations \cite{Jan} and Loop Quantum Gravity (LQG)
\cite{Rovelli:2004tv}  have
also drawn a lot of attention of theorists  from the mid 90's 
(see for instance \cite{Connes:1997cr}\cite{Doplicher:1994tu}\cite{Reuter:1996cp}\cite{Rovelli:2004tv})
when, in fact, the String's  revolution enchanted most of the physics community. 

In the meantime, another framework starts to build up 
around the Sakharov's idea (1965) of an ``emergent''  theory
of gravity (see \cite{Visser:2002ew} for a review). 
Mainly ``emergent'' refers to a phenomenon which is only
induced and not fundamental. The analogy with hydrodynamics
as emerging from laws of molecular physics is commonly used
as an illustration. 
This idea, somehow rooted in condensed matter physics and 
statistical physics, suggests only, but very originally, 
that the quantization of the spacetime background metric 
should be addressed in some independent way than 
the ordinary study of fluctuations around the flat metric configuration. 
In fact, it is well known that 
the latter type of quantization leads inexorably to a non renormalizable 
quantum field theory using Einstein-Hilbert action \cite{feynman, Goroff:1985sz}. 
So far, results on the QG domain are various and are greeted
with more or less success. Nevertheless, 
the idea of an emergent spacetime was welcome as pertinent
and still perpetuates through the years \cite{Konopka:2006hu}. 

Coming back to the mid 80's, the so-called random matrix models, 
initially motivated by string theoretical considerations, become
on their own a concrete proposal for statistical models for QG in 2$D$ \cite{Di Francesco:1993nw}.
Soon after, they were followed by their tensor analogue 
for $D>2$ QG \cite{ambj3dqg}-\cite{sasa1}. However,
the latter works experienced difficulties because 
one of the main analytical tool, namely 
the $1/N$ expansion, was crucially missing. In last resort, only 
numerical results can be properly achieved. All these endeavors
 appropriately realize the idea of an emergent theory of gravity
(in short, in $2D$, random matrix models are 
theories of random triangulations of surfaces 
and summing on such triangulations amounts to sum
over geometries of these surfaces. These models 
possess a phase transition towards a ``continuous phase'' geometry
the so-called conformal geometry coupled to Liouville
gravity \cite{Di Francesco:1993nw}). 
 
The framework of tensor fields for addressing the QG issue 
was correctly 
stated in \cite{Boul} for a $D$-dimensional lattice version of 3D Euclidean 
gravity. The prime idea was not so much focused on the 
understanding of an extended version of statistical analysis 
of matrix models for tensors but rather to introduce a
discrete version of a simplified and ``flat'' (BF) model of gravity valid in dimension higher  than 2. 
Substituting the integral over all $D$-dimensional manifolds by a
sum over all $D$-dimensional simplicial complexes,
the ensuing models generate perturbatively all $D$ simplicial
complexes providing the tensor analogues of the successful 
matrix models.   
Fields here are defined over several copies of an abstract Lie group
which, by Fourier transform, yield tensor components. 
Depending on the dimension, to each tensor is associated 
a simplex and vertices provide fusion rules and exchange of momenta
 as in ordinary quantum field theory. 
A key point of this formalism is that these tensors 
should be constrained to satisfy specific rules of invariance 
enforcing the flatness condition of the gluing of the simplices
at the vertices. A generic link was later found between these lattice models and another
fast-growing field: the spin foam models which embody the covariant version of the LQG program
 \cite{Freidel:1999jf, Reisenberger:2000zc}.
Thus, was born a new line of investigation for QG, the so-called group field theory (GFT) \cite{laurentgft,oriti}. 
The partition function of GFT sums over both topologies 
(dictated by the topology of Feynman graphs as simplicial manifolds) 
and geometries (encoded in the amplitude of these graphs)
of a given manifold. GFT claims to be a theory of quantization
``of'' spacetime itself \cite{oriti}. This leads, once again, 
to the problem of obtaining an emergent spacetime 
at some proper continuum limit.

The GFT framework appears very appealing for performing quantum
field theory computations and, naturally, the question of its
renormalizability was first systematically addressed \cite{FreiGurOriti,sefu1}. 
Many other interrogations arose concerning the GFT formalism.
``How to control divergences ?'' and  ``why the 
 most important contributions in the partition function
would be of the form of a ``large'' and ``smooth'' spacetime 
like the one, we experience ?'' were frequently asked.
In fact, all these interrogations could have been only satiated by 
finding a $1/N$ expansion for tensor models. 

The renormalization program for GFT made its first steps
and notable facts have been sorted out concerning power-counting
theorems of both ordinary GFTs and more involved GFT 
models in relationship with $4D$ QG \cite{sefu2}-\cite{BS1}. 
In the meantime, a drastic improvement concerning the topology of  simplices dual to the Feynman graphs has been proved by Gurau by introducing the colored version of these GFTs \cite{color}\cite{Gurau:2010nd}. In \cite{Geloun:2011cy},
it appears clear that relevant operators of the Laplacian form
\cite{Oriti:2010hg} were missing in the $3D$ Boulatov GFT action \cite{Boul} and, so, should be added in all GFT actions before discussing of their renormalizability. 
The renormalization analysis for these type of GFTs proves 
to be a computational challenge. 
Concerning the symmetry analysis of 
GFTs, some efforts have been made in order to show
that they possess a quantum group symmetry \cite{Baratin:2011tg},
that they satisfy peculiar Ward identities \cite{BenGeloun:2011xu} 
and that, associated with translation invariance, 
there exists a conserved classical energy momentum tensor  for the colored GFTs 
\cite{BenGeloun:2011cz}.

The real breakthrough of the story occurs with the major
discovery by Gurau of the analogue of $1/N$ expansion  
both for GFTs and independent identically distributed tensor models
(see \cite{Gur3}\cite{Gur4} and more other references therein, 
 \cite{Rivasseau:2011xg} for a short review of the subject
and \cite{Gurau:2011xp} for a complete overview of the subject and more
developments based on this $1/N$ expansion). 
Dominant graphs were identified \cite{Bonzom:2011zz}
and are associated with simplices of the sphere topology. Important developments followed:
the generalization of the Witt-Virasoro algebra (without central 
charge) for infinite dimensional tree algebras \cite{Gurau:2011tj}, 
the tensor generalization of the Ising model in any dimension 
\cite{Bonzom:2011ev}, the determination of the 
universal character of random tensor models generalizing the Wigner-Dyson law for random matrices \cite{Gurau:2011kk}. 
Furthermore, based on the above developments and for the first time,
a $4D$ QG model, even though simplified,  
has been found renormalizable at all order of perturbation 
theory \cite{arXiv:1111.4997}. Let us readily mention that 
it is not clear in which sense General Relativity could be
recovered from the said model (this is why one calls it simplified). However, it is not hopeless to encounter
coordinate invariance and more geometrical properties of 
the background space emerging  through the rich structure of that model 
 \cite{Rivasseau:2011hm}. 

The idea of defining a renormalizable and emergent theory of gravity using the tensor approach has matured through years  \cite{Rivasseau:2011hm} and belongs to a particular pattern extending progressively the renormalization group (RG) analysis from local graphs like in the $\phi^4_4$ model, to the vector
cases for condensed matter systems, then to the matrix cases like in noncommutative field theory \cite{Rivasseau:2007ab}-\cite{Gurau:2008vd} incorporating already nonlocality and, now, to higher dimensional tensor theories. The model 
of \cite{arXiv:1111.4997} is nonlocal and defined over 
copies of the compact group $U(1)$ hence does not have infrared (IR) divergences. 
It is built by first integrating four colors (out of five) 
in the colored model \cite{color} and choosing the last color to be dynamical with respect to \cite{Geloun:2011cy}. 
Considering the $1/N$ expansion, it is known that only certain 
contributions in the effective action called ``melonic'' (this terminology refers to \cite{Bonzom:2011zz}) will be not suppressed in power-counting. 
 It remains to truncate the effective action 
for obtaining relevant and marginal operators of the melonic
kind.    
After analyzing the divergence degree of an arbitrary graph 
at a given order of perturbation theory, 
only graphs determined by  particular boundary data but 
having also a regular internal structure turn out to be divergent.
 This is the ``generalized locality'' principle
for tensor graphs and only those graphs should be renormalized by standard interpolation moves on external data. 
Interestingly, a peculiar anomaly arises in the expansion.
The authors of \cite{arXiv:1111.4997} 
interpreted this term as matter defects appearing in that $4D$ gravity model. 
Renormalization group (RG) flows have been not computed for the model and these deserve indeed to be understood. As it can be easily realized, 
the combinatorics of such a model becoming very involved should be cautiously handled. The present contribution 
is a step towards the computation of the RG flow of that $4D$ 
QG. 

In this paper, we address the RG flows of coupling constants of a 
tensor model in $3D$ related to the above $4D$ model. This model has been briefly outlined 
in \cite{arXiv:1111.4997} but its renormalizability has been 
not yet proved. This is an initial property that we need to investigate
before computing any flow.
We prove that this $3D$ tensor model over $U(1)^3$
is renormalizable at all orders of perturbation. The proof
is thoroughly done in the momentum space.  This is in contrast 
with the renormalizability proof of the anterior $4D$ model 
which was performed in the direct space. 
In the momentum basis, the interaction remain of the same form
of as in the direct space yielding factorized graph amplitudes 
somehow more suitable to perform 
the different optimizations occurring in the multiscale analysis. 
Hence, using a different basis also convey to a new perspective
on these tensor models in general.
In a second part of this paper, we compute the one-loop $\gamma$- 
(governing the RG flow of the wave function renormalization) 
and $\beta$-functions (governing the RG flow of interaction 
coupling constants) and analyze the RG flows of the different coupling
constants. We emphasize that it is not necessary to go beyond 
 one-loop computations since the leading order corrections
determine the RG flows of coupling constants 
if the $\beta$-function is not vanishing at this order. 
The model obtained by merging all coupling constants
to a unique one which is somehow the most natural model
proves to be asymptotically free in the UV. 
This feature might be related to the universal Gaussian 
behavior of these tensor models discussed in \cite{Gurau:2011kk}.

The plan of this paper is as follows: the next section 
defines the model and states our two main results. 
Section 3 is devoted to the multiscale analysis 
and the achievement of a crude power-counting theorem
which will be dissected in Section 4. 
The renormalization of primitively divergent graphs 
will be performed in Section 5. The calculations 
of $\gamma$- and $\beta$-functions and RG flows
of coupling constants are detailed in Section 6. 
Section 7 gives a summary of our results and 
an outlook of this work.
An appendix gathers further details on some 
results used  through the text.

\section{Rank  3 tensor model over U(1)}
\label{sect:model}

Let us consider four complex three rank tensor fields over the group $U(1)$, $\varphi^{\rm a}:U(1)^3\longrightarrow\mathbb{C}$.
The index ${\rm a}=0,1,2,3$ is called color. 
In Fourier modes, $\varphi^{\rm a}$ can be expanded as 
\beq
\varphi^{\rm a}(g_1, g_2, g_3)=\sum_{p_j\in\mathbb{Z}}\varphi^{\rm a}_{[p_j]}e^{ip_1\theta_1}e^{ip_2\theta_2}e^{ip_3\theta_3},\quad \theta_i\in [0,2\pi)\quad \mbox{ and }\quad [p_j]=(p_1,p_2,p_3),
\eeq
where the group elements $g_k=e^{i\theta_k}\in U(1)\cong S^1$.
From now on, we write $ \varphi^{\rm a}_{123}:= \varphi^{\rm a}_{p_1,p_2,p_3}$
and define a theory in the momentum space with a first kinetic part 
regarding the colors ${\rm a}=1,2$ and $3$, 
\beq
S^{\kin;1,2,3}=\sum_{p_j} \sum_{{\rm a}=1}^3 \bar{\varphi}^{\rm a}_{123}\,\varphi^{\rm a}_{123},
\eeq
where the sum in $p_j$ is performed over all momenta values, for $j=1,2,3$.
We do not assume any symmetry under permutation of arguments of these tensors. 

The ordinary colored theory \cite{color} is defined by an interaction which can be
read off in momentum space as 
\beq
S^{\inter}= \tilde\lambda
\sum_{p_j}
\varphi^0_{123} \,
\varphi^1_{345} \,
\varphi^2_{526} \,
\varphi^3_{641}\,
+\bar{\tilde{\lambda}}
\sum_{p_j}
\bar\varphi^0_{123}\,
\bar\varphi^1_{345}\,
\bar\varphi^2_{526}\,
\bar\varphi^3_{641},
\label{vertexcolo}
\eeq
the parameters $\tilde\lambda$ and $\bar{\tilde{\lambda}}$ being the coupling constants. 

We emphasized here an important point concerning these
tensor models. The momentum space is 
in ``exact duality'' with the direct space in the following sense:
up to an inessential constant (a power of $2\pi$ coming from
spatial integrations), interactions in direct and momentum spaces
share exactly the same form. This is contrast with other local
theories or even nonlocal theory like in noncommutative field theory \cite{Gurau:2008vd} for which the vertex possesses a delta
function of momentum conservation. Although 
nothing prevents to perform the renormalization in the direct space,
the main point for switching in momentum space is that graph amplitudes 
get factorized in a  different and useful way as we will see.

The polar point here is to take a particular kinetic term with respect to the
color $0$ such that
\beq
S^{\kin,0} =\sum_{p_j} \bar{\varphi}^0_{123}\Big(\sum_{s=1}^3 a_s|p_s| + m\Big)\varphi^0_{132}, 
\label{eq:action0}
\eeq
where $|p_s|$ denotes the absolute value of $p_s$ and
$a_s$ are positive wave-function coupling constants.
Hence the field of color $0$ is assumed to be propagating.\footnote{
Remark that a direct space formulation corresponding to such a 
momentum space kinetic term can be defined by a reduced operator
acting on each strand as 
 $-i \tilde\partial_{\theta}\phi(\theta) = \sum_{p \in \Z} |p| \tilde \phi(p) e^{i p \theta}$, where $\tilde \phi(p)$ is the Fourier
mode of $\phi$.} Note that also, the model described by 
\eqref{eq:action0} is  slightly different from the direct 3D analogue
of \cite{arXiv:1111.4997}. The latter could be only reproduced
by fixing all $a_s$ to a given value.

\begin{figure}
 \centering
     \begin{minipage}[t]{.8\textwidth}
      \centering
\includegraphics[angle=0, width=3.5cm, height=0.9cm]{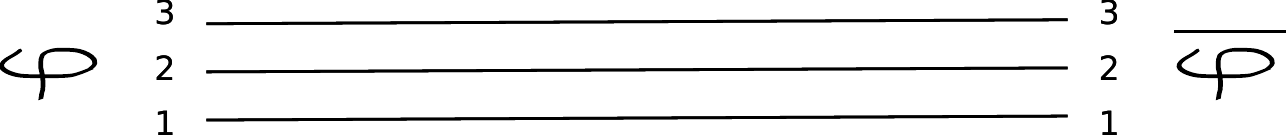}
\caption{ {\small Propagator. }}
\label{fig:propa}
\end{minipage}
\end{figure}

\begin{figure}
 \centering
     \begin{minipage}[t]{.8\textwidth}
      \centering
\includegraphics[angle=0, width=14cm, height=4.5cm]{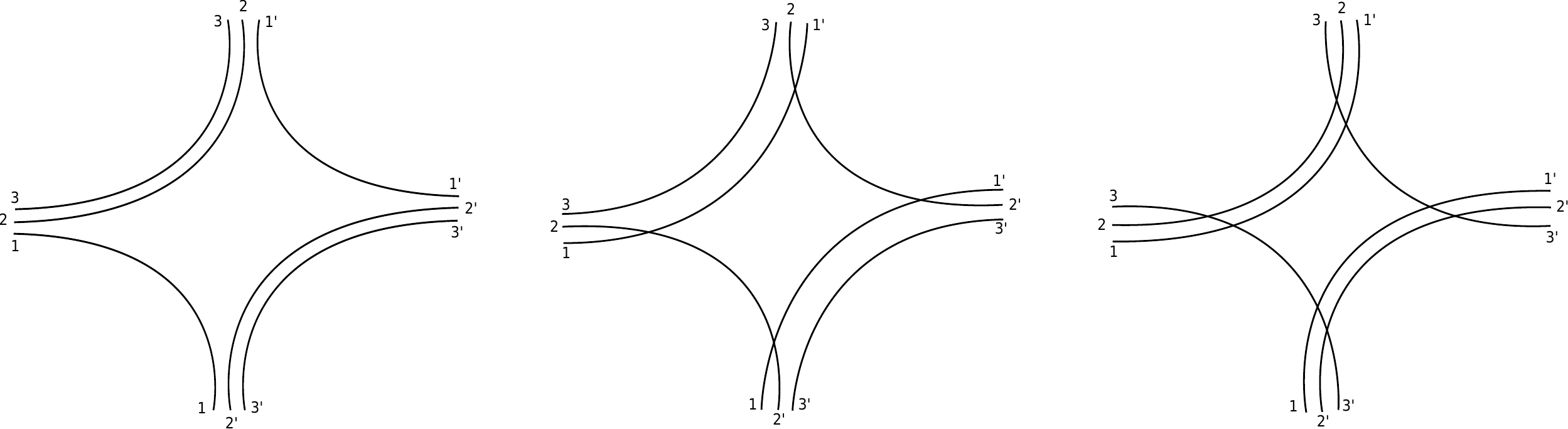}
\caption{ {\small Vertices of the type $V_{4}$. }}
\label{fig:vertex}
\end{minipage}
\end{figure}

We use the same procedure of  \cite{Gurau:2011tj} which mainly 
performs an integration of
the partition function with respect to all  colors save one. 
One gets an effective action for the last tensor $\varphi^0$ in the form
\beq
\mathcal Z = \int d\mu_C[\varphi^0]\; e^{- S^{\inter,0}  },  
\eeq
where $d\mu_C[\varphi^0]$ stands for the Gaussian measure with covariance 
$C=(\sum_s a_s|p_s|+m)^{-1}$  (represented in Fig.\ref{fig:propa}) and the effective interaction finds the form 
\beq
\label{integrateco}
S^{\inter, 0} =
\sum_{\mathcal B}
\frac{(\tilde\lambda\tilde{\bar\lambda})_{\mathcal B} }{
\text{Sym}(\mathcal B)} N^{f(p,D) -\frac{2}{(D-2)!}\omega(\mathcal B)}
\text{Tr}_{\mathcal B} [\bar\varphi^0\varphi^0];
\eeq
the sum in $\mathcal B$ is performed on all ``bubbles''
which are connected vacuum graphs with colors $1$ up to $D=3$
 and $p$ vertices; $f(p,D)$ is a function of the number of
vertices and the dimension; $\omega(\mathcal B) := \sum_{J} g_{J}$
is the sum of genera of sub-ribbon graphs called jackets $J$
of the bubble, and
$\text{Tr}_{\mathcal B}[\bar\varphi^0\varphi^0]$ are called
tensor network operators.
Graphs with $\omega(\mathcal B) =0$ are called
melons \cite{Gurau:2011xp} and non melonic contributions defined
by $\omega(\mathcal B) >0$ become suppressed from (\ref{integrateco}).
Thus, it is sufficient to only focus on the melonic sector of the theory. 
As in  \cite{Gurau:2011tj}, one attributes a different coupling constant to different  tensor network operators and
simply rewrites (omitting the color index $0$)
\beq
S^{\inter, 0} =
\sum_{\mathcal B}
\frac{\lambda_{\mathcal B}}{
\text{Sym}(\mathcal B)}
\text{Tr}_{\mathcal B} [\bar\varphi \varphi ].
\eeq
We truncate the above series  and will consider only relevant to marginal terms guided by renormalization conditions. 
Thus, in the following, we will consider the effective interaction terms made of 
monomials with order at most four: 
\beq
S_{4}=\sum_{p_j}\varphi_{p_1,p_2,p_3}\,\bar{\varphi}_{p_{1'},p_2,p_3}\,
\varphi_{p_{1'},p_{2'},p_{3'}}\,\bar{\varphi}_{p_{1},p_{2'},p_{3'}}
+\mbox{permutations}.
\label{vertex4}
\eeq
The last term ``permutations'' means that we include also other terms by performing
a permutation over the six momentum arguments (see 
Fig. \ref{fig:vertex}).

By introducing the UV cutoff $\Lambda$ on the propagator which becomes $C^\Lambda$, the action and  partition function of our model are then defined by
\begin{eqnarray}
S^\Lambda=   \lambda^{\Lambda}_{4}  S_{4}+ C T_{2;1}^\Lambda\, S_{2;1}+\sum_{s=1}^3 C T_{s;2;2}^\Lambda\, S_{s;2;2},\qquad 
\mathcal Z = \int d\mu_C^\Lambda[\varphi]\; e^{- S^{\Lambda}  },
\label{actioncut}
\end{eqnarray}
where
\begin{eqnarray}
S_{2;1}=\sum_{p_j}\bar{\varphi}_{[p_j]}\varphi_{[p_j]},\qquad S_{s;2;2}=\sum_{p_j}\bar{\varphi}_{[p_j]}\Big(|p_s|\Big)\varphi_{[p_j]},
\end{eqnarray}
where the symbols $CT$'s are coupling constant counterterms defined
by the difference between the renormalized and bare couplings. 
More precisely, $C T_{2;1}^\Lambda$ is a mass counterterm
and $C T_{s;2;2}^\Lambda$ is a wave function counterterm.

The main results of this paper are given by the following
statements:

\begin{theorem}
\label{theo:ren}
The model defined by \eqref{actioncut} is renormalizable at
all orders of perturbation theory.
\end{theorem}

\begin{theorem}
\label{theo:asfree}
The model obtained from  \eqref{actioncut}  
by identifying $a_s = a$ is asymptotically free in the UV direction.
\end{theorem}

\section{Multiscale analysis}
\label{sect:multi}

We begin with the multiscale analysis which will lead to a 
prime (or crude) power-counting
theorem. From that theorem, we will be able to identify the ``dangerous'' 
graphs for which the renormalization program should be performed.
The first step is to find the behavior of the propagator  with respect to 
high and low momentum scales, then, using this result, we should find an optimal way to bound
graph amplitudes in the most general manner.

\subsection{Propagator bound}

Consider the kinetic term of the action. The kernel of the propagator is
\begin{eqnarray}
C([p_s],[p'_s])=\Big[\sum_{s=1}^3a_s|p_s|+m\Big]^{-1}\Big[\prod_{s=1}^{3}\delta_{p_s,p'_s}\Big],
\end{eqnarray} 
the notation $C([p_s],[p'_s])$ referring to $C([p_1,p_2,p_3],[p'_1,p'_2,p'_3])$.

Introducing a Schwinger parameter, we get the integral form of the propagator as
\beq
C([p_s],[p'_s])=\int_0^\infty \, d\alpha \,e^{-\alpha[\sum_{s=1}^3a_s|p_s|+m]}.
\label{cpp}
\eeq
In the following developments, only is needed 
the small distance or high momenta behavior of that propagator.
We do not have the infrared divergence problem from 
the compactness of $U(1)$. This is reflected in the direct space (Fourier transform of the above), for instance, by the fact that the mass can be put to $m= 0$ and this will cause no difficulty  with the zero mode of the propagator. Working in the momentum space, in full generality, 
it is better consider a non vanishing mass otherwise low momenta 
will cause a divergence in \eqref{cpp}. Nevertheless, as mentioned earlier,  this situation is actually not relevant for the remaining analysis
which solely focuses on the UV sector of the theory.

The next stage is to introduce the slice decomposition of the propagator in the form:
\begin{eqnarray}\label{scaledec}
&&
C  = \sum_{i=0}^{\infty} C_i \;\quad i=1,2\cdots \\
&&
C_0=\int_1^\infty \, d\alpha \,e^{-\alpha[\sum_{s=1}^3a_s|p_s|+m]},\quad 
C_i=\int_{M^{-i}}^{M ^{-i+1}}\, d\alpha \,e^{-\alpha[\sum_{s=1}^3a_s|p_s|+m]},\, \quad M \in \mathbb{N} .
\end{eqnarray}
The following proposition is direct
\begin{lemma}
For all $i\in\mathbb{N}$,  there exists a large constant $K\geq 0$  such that
\beq\label{boundprop}
C_i  \leq KM^{-i}e^{- M^{-i}|\sum_{s=1}^3 a_s|p_s|+m|}.
\eeq 
\end{lemma}
The ultraviolet cutoff can be imposed by summing the slice index $i$ up to a large integer
called $\Lambda$ in  the slice decomposition \eqref{scaledec}. Thus
\beq
C^{\Lambda}  = \sum_{i=0}^{\Lambda} C_i ,
\eeq
and the ultraviolet limit consists in taking $\Lambda\to\infty.$ 
 One calls $C_0$ and $C_\Lambda$ the IR and UV propagator slice, respectively. 
For simplicity in the following, we omit the superscript $\Lambda$.

\subsection{Optimal amplitude bound: Prime power-counting}
\label{subsect:ppc}

Let $\mathcal{G}$ be  a  connected amputated graph with set of vertices $\mathcal{V}$  (of any kind for the moment) of cardinal $V$ and $\mathcal{L}$  set of lines of cardinal $L=|\mathcal{L}|$.
The bare amplitude associated with $\cG$ is of the form
 \bea
A_{\cG} &=&\sum_{\mu}\frac{f_\mu[\lambda,CT_{2,1},CT_{s;2;2}]}{S(\cG)}A_{\cG,\mu}\crcr
A_{\cG,\mu} &=& \sum_{p_{v,s}} 
\Big(\prod_{\ell \in \cL} C_{i_\ell(\mu)}
([p_{v(\ell), s}];[p_{v'(\ell), s}])\Big)
\prod_{v\in \cV; \;s}  \delta_{p_{v, s},\,p_{v', s}} ,
\eea 
where  $f_\mu[\lambda,CT_{2,1},CT_{s;2;2}]$ is a function of some 
product of coupling parameters, $S(\cG)$ is a symmetry factor,  $p_{v(\ell), s}$ are momenta involved in the propagator which 
should possess a vertex label $v(\ell)$ hooked to a given line $\ell$
and a strand label $s$; $p_{v,s}$ are 
the same momenta, but now, involved in the vertex
which should bear both vertex $v$ and strand $s$ indices;
$\delta_{p_{v, s},p_{v', s}}$ is the Kronecker symbol  (we keep 
here formal notations but, while dealing with a category of vertex $V_4$, $V_2$, this symbol might depend on the vertex type and
the number of strands $s$ could vary from one category of vertices to the other; however, as we will see below, after some momentum summations, we do not need to 
track all these indices to get a crude power counting);
$\mu= (i_1,i_2,\dots,i_q)$ is called momentum 
assignment and gives to each propagator of each internal line 
$\ell$  a scale $i_\ell  \in [0,  \Lambda ]$; the sum over $\mu$ is performed on all assignments.  From the point of view
of the effective series expansion \cite{Rivasseau:1991ub}, the function $f_\mu$ collects all the effective couplings corresponding to
the attribution $\mu$.
Given an amputated graph, 
we simply have external vertices (where test 
functions or external fields can be hooked). 
By convention, we fix all external line scales at $i_{\ext}=-1$. 
$A_{\cG;\mu}$ will be the core quantity  
and the sum  $A_{\cG}=\sum_{\mu} [f_\mu(\lambda,CT_{\dots})/S(\cG)]A_{\cG;\mu}$
can be done only after renormalization in the standard way of \cite{Rivasseau:1991ub}.

We would like to perform the momentum sums in the $p_{v,s}$ 
in an ``optimal'' way. For that purpose, let us quickly review  
what should be expected from the ordinary theory. 
Given $\mu$ and a scale $i$, we consider the complete list
of the connected components $G_i^k$, $k=1,2,\dots,k(i)$, 
of the subgraph $\cG_i$ made of all lines in $\cG$
with the scale attribution $j\geq i$ in $\mu$ 
 (note also that the meaning of $i$ and $k$ are radically different
even though we keep simple notations for these quantities
when writing $G_i^k$). 
Such subgraphs are called high or quasi-local 
and are the key objects in the multiscale expansion  \cite{Rivasseau:1991ub}.
There is a partial (inclusion) order on the set of $G^k_i$ and $\cG_0=\cG$. The abstract tree 
made of nodes as the $G^k_i$ associated to that partial order
is called the Gallavotti-Nicol\`o tree \cite{Galla}. 
$\cG$ is the root of that tree. To an arbitrary subgraph $g$, 
one assigns two quantities:
\beq
i_g(\mu) = \inf_{l\in g}i_l(\mu) \;,  \qquad e_g(\mu) = 
\sup_{l \,\text{external line of}\, g} i_l(\mu) .
\eeq
A subgraph $g$ can be viewed as $G_i^k$ for a given $\mu$ 
if and only if $i_g(\mu) \ge i >e_g(\mu) $.
In the direct space formalism, one considers a spanning tree\footnote{This 
is by definition a subgraph formed by lines passing through all vertices without
forming closed loops.} $T$
of the graph $\cG$. Associated with $T$ are position variables
that one  integrates to give decay factors to the product of propagator lines. 
The key point is to optimize the bound over spatial integrations 
by choosing the tree $T$ to be compatible with the 
Gallavotti-Nicol\`o tree. This is achieved by taking
the restriction $T_i^k$ of $T$ to any $G_i^k$  such 
that $T_i^k$ is  again a spanning tree for $G_i^k$. 

\begin{figure}
 \centering
     \begin{minipage}[t]{.8\textwidth}
      \centering
\includegraphics[angle=0, width=15cm, height=6cm]{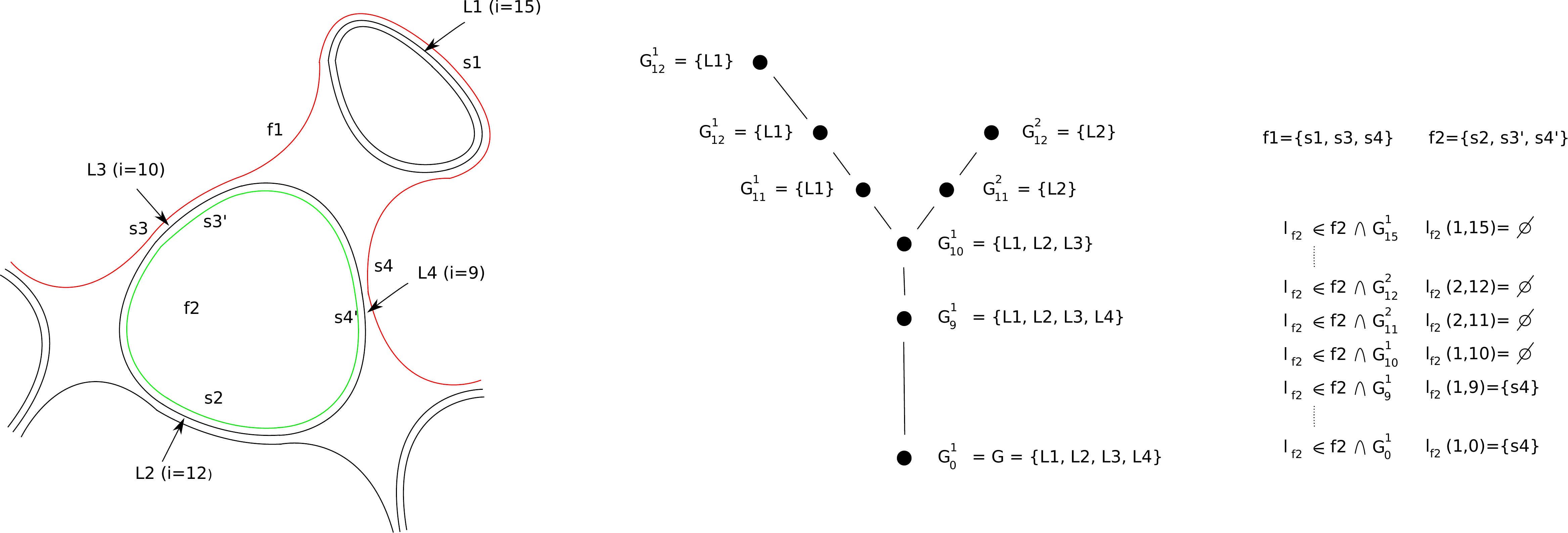}
\caption{ {\small A graph with a multiscale expansion ($L1$ is at scale $i=15$, etc.) and
 its Gallavotti-Nicol\`o tree; the face $f1$ (in red) is external 
and the face $f2$ (in green) is internal. 
 $l_f$ is the strand element  at given scale $i$ which 
should optimize the bound.}}
\label{fig:gtrees}
\end{minipage}
\end{figure}

Coming back to our situation and for simplicity,  we assume that no wave function counterterm appears in the graph. Adding them at the end will be an easy task. One notices that the vertex operator
is a product of delta functions and hence $A_{\cG}$ factorizes
in term of closed or open strand line that we call ``faces''. 
Let $\cF$ be the set of faces of $\cG$. 
It decomposes in a set $\cF_{\inter}$ of closed (or internal) faces
and another set $\cF_{\ext}$ of faces connected to external vertices. 
We have 
\beq
|\cF| = F = F_{\inter} + F_{\ext}, \qquad  F_{\inter} = |\cF_{\inter}|, \qquad  
F_{\ext} = |\cF_{\ext}| ,
\eeq
and
\bea
|A_{\cG;\mu}|
&\leq&  K'^n 
\prod_{\ell \in \cL}  M^{-i_\ell } 
\sum_{q_{s}}
\prod_{\ell \in \cL} [\prod_{s=1}^3 \delta_{q_{i_\ell s}, q'_{i_\ell s}}]
 \,e^{- M^{-i_\ell}[\sum_{s} a_s|q_{s}| + m]} \crcr
&\leq&
 K'^n 
\prod_{\ell \in \cL}  M^{-i_\ell } 
\sum_{q_{f}} 
\prod_{f \in \cF} \prod_{\ell \in f}
e^{-  M^{-i_\ell}\, a_{f}|q_{f}| } ,
\label{boundpresq}
\eea
where $a_f |q_f|$ is the wave function quantity now bearing a face
label $f$. The bound (\ref{boundprop}) has been also used 
in order to get \eqref{boundpresq}. 
The following cases may occur:
\begin{enumerate}
\item[(i)] The face $f \in \cF_{\inter}$, then the face amplitude is 
$\sum_{q_f} e^{-(\sum_{\ell \in f} M^{-i_\ell})a_f |q_{f}|}$.  
 Given $i=\min_{\ell\in f}i_\ell$, 
this amplitude can be optimized as, up to some constant $\delta$, 
\bea
\sum_{q_f} e^{-(\sum_{\ell \in f} M^{-i_\ell})a_f |q_{f}|} 
\leq \sum_{q} e^{- M^{-i} a_f |q|} = \delta M^{i} + O(M^{-i}) .
\eea
\item[(ii)] the face $f$ is open, then all sums in $q_s$
can be performed and yield  $O(1)$. 
\end{enumerate}
Hence, in the above amplitude \eqref{boundpresq},
only terms involving $\cF_{\inter}$ should be taken into account. We obtain
\bea
|A_{\cG;\mu}|
&\leq&
 K'^n 
\prod_{\ell \in \cL}  M^{-i_\ell } \sum_{q_{f}}
\prod_{f \in\cF_{\inter}} 
e^{-(\sum_{\ell \in f} M^{-i_\ell})\,a_f |q_{f}| } \crcr
&\leq& K'^n\prod_{\ell \in \cL}\prod_{i=1}^{i_\ell}M^{-1}\prod_{f \in\cF_{\inter}}K''M^{i_{l_f}},
\eea 
where $l_f$ is a strand of $f$ such that $i_{l_f}= \inf_{l\in f} i_l$. 
It can be therefore inferred that
\bea
|A_{\cG;\mu}|
&\leq& K'^{n}K''^{F_{\inter}}\prod_{\ell \in \cL}\prod_{i=1}^{i_\ell}M^{-1}\prod_{f \in \cF_{\inter}}\prod_{i=1}^{i_{l_f}}M\cr
&\leq& K'^{n}K''^{F_{\inter}} \prod_{\ell \in \cL}\prod_{(i,k)\in \mathbb{N}^2 / \ell \in G_i^k} M^{-1} \prod_{ f \in \cF_{\inter}} \prod_{(i,k)\in \mathbb{N}^2 / l_f \in G_i^k} M\cr
&\leq& K'^{n} K''^{ F_{\inter}}
\prod_{(i,k)\in \mathbb{N}^2} M^{ - L(G^k_i)}\prod_{(i,k)\in \mathbb{N}^2}
\prod_{f \in \cF_{\inter} \cap G_i^k} \prod_{l_f \in f \cap  G_i^k }M\cr
&\leq&K'^{n} K''^{ F_{\inter}}\prod_{(i,k)\in \mathbb{N}^2}M^{- L(G^k_i) + F_{\inter}(G^k_i)},
\eea
 where $ L(G^k_i)$ and $F_{\inter}(G^k_i)$ denote
the number of internal lines and internal closed faces of $G^k_i$.
In the last step,  during the bound optimization, one uses the fact that $l_f \in f \cap  G_i^k$ gives a single  strand the scale of which is the minimum among the scales 
of the lines which occur in $ f \cap  G_i^k$.
Note that the face $ f \cap  G_i^k$ can be an open face. In such a case,
 $l_f$ being the minimum of the scales of the closed face $f$ cannot occur in 
the $ f \cap  G_i^k$, hence the product in $l_f \in f \cap  G_i^k$ is empty. 
 Thus, the last product yields nothing but the number of elements of the internal 
closed faces of $G_i^k$ (when the closed face $f$ becomes totally embedded in the $G_i^k$ such that $l_f  \in G_i^k$;
 see Fig.\ref{fig:gtrees}).

Compounding all constant factors coming from the 
momenta sums, we get a bound of the
graph amplitude at a given attribution $\mu$ as
\beq  \label{goodbound}
|A_{\cG;\mu}| \leq K^{n} \prod_{(i,k) }
M^{- L(G^k_i) + F_{\inter}(G^k_i)} ,
\eeq
where $K$ is some constant and $n$ is the number of vertices of the graph.
We recall that the above amplitude is assumed to be 
without wave-function counterterms. Adding these counterterms
in the form of $V'_{s;2}$ vertices, the following statement is straightforward:
\begin{lemma}[Prime power-counting]
\label{primepow}
For a connected graph $\cG$ (with external arguments integrated
versus fixed smooth test functions), we have
\beq
|A_{\cG;\mu}| \leq K^n  \prod_{(i,k)}
M^{ \omega_d(G^k_i)},
\eeq
where $K$ and $n$ are large constants, 
 \beq
\omega_d(G^k_i)= - L(G^k_i) + F_{\inter}(G^k_i)  +  \sum_{s=1}^3V'_{s;2} (G^k_i)  .
\eeq 
\end{lemma}
The quantity $\omega_d(\cG)$ is called the divergence of the graph $\cG$
and provides also the power-counting. In the sequel, we aim at analyzing this quantity
for a general connected graph $\cG$.

\section{Analysis of the divergence degree}
\label{sect:cdegan}

The analysis of the divergence degree is made in two steps:
the first by translating  $\omega_d(\cG)$ in topological
quantities and the second by refining the obtained result  
in a convenient manner. From these two steps, we provide
 an exhaustive list of divergent graphs. 

\subsection{Power-counting in topological terms}
\label{subsect:pctopo}

\begin{figure}
 \centering
     \begin{minipage}[t]{.8\textwidth}
      \centering
\includegraphics[angle=0, width=10cm, height=2cm]{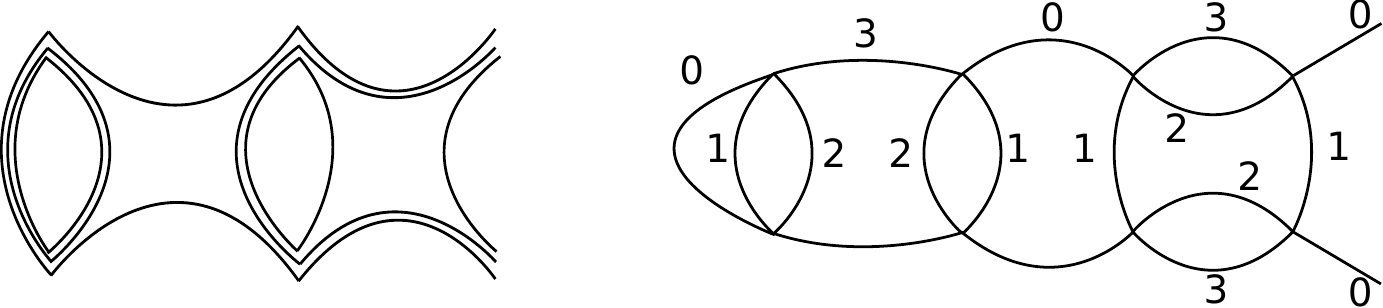}
\caption{ {\small { A graph $\cG$ (left) and its color extension} $\cexG$
(right) in simplified notations: each line of color $\alpha=0,1,2,3$ 
corresponds to a propagator in the colored theory, i.e. $\int d\mu_C(\varphi) \bar\varphi^\alpha_{123}\varphi^\alpha_{123}$,
and vertices are defined by Eq.\eqref{vertexcolo}.}}
\label{fig:colorext}
\end{minipage}
\end{figure}

\begin{figure}
 \centering
     \begin{minipage}[t]{.8\textwidth}
      \centering
\includegraphics[angle=0, width=12cm, height=2cm]{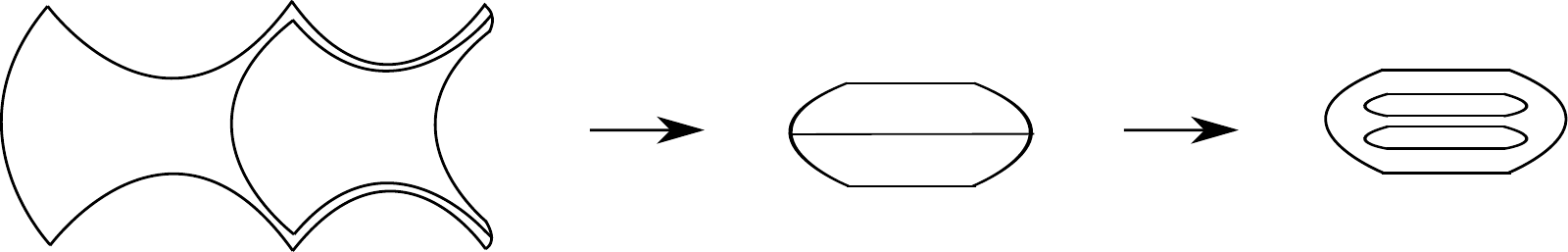}
\caption{ {\small  { The boundary $\bG$ of $\cG$ and its 
 rank two or ribbon structure.}}}
\label{fig:boundary}
\end{minipage}
\put(-310,-12){$\bG$}
\put(-165,-12){$\bG$}
\put(-58,-12){$\bG=J$}
\end{figure}

We recall some definitions (see \cite{Bonzom:2011zz}\cite{Gurau:2009tz}\cite{arXiv:1111.4997}):
\begin{definition}
Consider $\cG$ a $3$ dimensional graph. 
\begin{enumerate}
\item[(i)]  The colored extension of $\cG$ is the unique  graph $\cexG$
obtained after restoring in $\cG$ the former colored theory graph
 (Definition 1 in \cite{arXiv:1111.4997}, here illustrated in 
dimension 3 in Fig.\ref{fig:colorext}).

\item[(ii)]  A jacket $J$ of $\cexG$ is a ribbon subgraph of 
$\cexG$ defined by a cycle $(0abc)$ up to a cyclic permutation  
 (see 
Definition 1 of \cite{Bonzom:2011zz} and an illustration given by Fig.\ref{fig:jacket}).  There are 3 such jackets due to the dimension 3.

\begin{figure}
 \centering
     \begin{minipage}[t]{.8\textwidth}
      \centering
\includegraphics[angle=0, width=12cm, height=2cm]{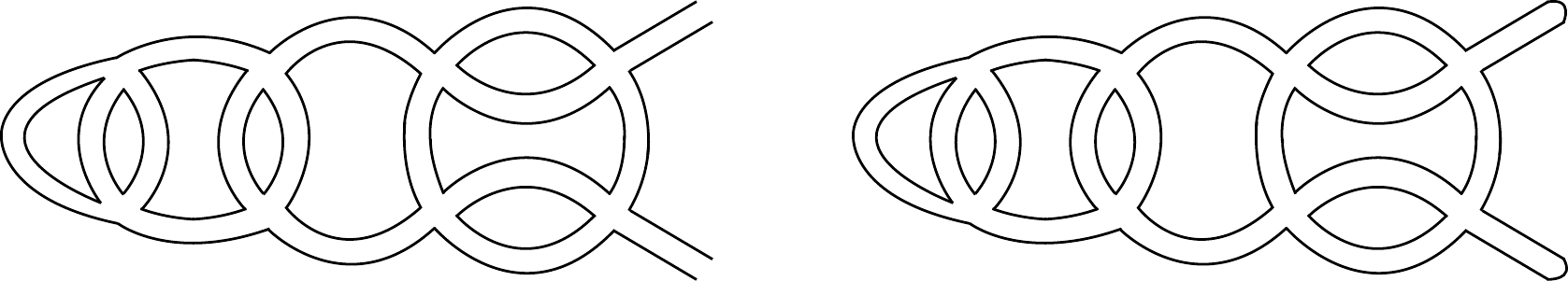}
\caption{ {\small The jacket $J$ (0123), ribbon subgraph of 
the colored graph $\cexG$ (Fig.\ref{fig:colorext}) and
the pinched jacket $\tJ$ associated with $J$.}}
\label{fig:jacket}
\end{minipage}
\put(-290,-12){$J$}
\put(-100,-12){$\tJ$}
\end{figure}

\item[(iii)] The jacket $\tJ$ is the jacket graph obtained from 
$J$ after ``pinching'' that is the procedure consisting in 
closing all external legs present in $J$ (see Section 3.2 
of \cite{Gurau:2009tz}, for the general definition of ``pinching'' 
for external strands; here the pinching of a jacket is
illustrated in Fig.\ref{fig:jacket}). 
It is always a vacuum graph. 

\item[(iv)]  The boundary $\bG$ of the graph $\cG$
is the closed graph defined by vertices corresponding to 
external legs and by lines corresponding to external strands of $\cG$ 
 (see Section 3.2 of \cite{Gurau:2009tz} as well as a drawing 
in Fig.\ref{fig:boundary}). Here, it is a vacuum graph
of the $2$ dimensional colored theory, hence a ribbon graph. 
It is also its own and unique jacket.

\end{enumerate}
\end{definition}

Some comments are in order. In general, the boundary graph 
of a tensor graph of rank $D$ is a closed tensor graph or rank $D-1$. The notion of jacket for the boundary graph can be defined as in the ordinary situation as a cycle of colors. The fact that 
 the boundary graph is defined as a closed graph induces immediately 
that all jackets are closed.

Let $\cG$ be a graph. Let $V_4$ be its number of $\varphi^4$
vertices, $V_{2}$ the number of vertices corresponding to mass-counterterms $\varphi^2$,
$V'_{s;2}$ the number of vertices corresponding to wave function counterterms
$|p_s| \varphi^2$. The graph possesses also $L$ number of lines and $N_{\ext}$ external 
fields.
Using the above notations, the following statement holds:
\begin{theorem}\label{theodeg}
The divergence degree of connected graph $\cG$ is given by
\beq
\omega_d(\cG)=
 -V_2  - \frac12  ( N_{\ext} -4)  - \sum_{J} g_{\tJ}  + g_{\bG}    -  (C_{\bG}-1),
\label{omeg}
\eeq
where the sum is performed on all  jackets $J$ of $\cexG$,
$g_{\tilde{J}}$ is  the genus of the jacket $\tilde{J}$ associated with 
$J$, 
$g_{\bG}$ is the genus of $\bG$ and $C_{\bG}$ is the number
of connected components of the boundary graph $\bG$.
\end{theorem}
\noindent{\bf Proof.}
There is a relation between the numbers of lines, of external legs and of vertices 
 for $\cG$: $4 V_4  +2 (V_2  + \sum_{s=1}^3V'_{s;2} ) = 2L + N_{\ext}$. Concerning $\cexG$, its number 
of vertices $V_{\cexG}$ and  number of lines $L_{\cexG}$ satisfy
\beq
V_{\cexG} = 4 V_4  + 2 (V_2  + \sum_{s=1}^3V'_{s;2})  , \quad
L_{\cexG} =  L + L_{\inter;\,\cexG} = \frac12 (4 V_{\cexG} - N_{\ext}),
\label{vexlex}
\eeq
where $L_{\inter;\cexG}$ denotes the number of internal lines of  $\cexG$ which 
do not appear in $\cG$.
 $F_{\cexG}$, the number of faces of  $\cexG$, 
can be partitioned in the number $F$ of
faces of the initial graph but also in additional
faces $F_{\inter;\cexG}$ proper to the colored graph 
and not appearing in $\cG$. We write
\beq
F_{\cexG} = F + F_{\inter;\,\cexG} .
\eeq
There are 3 jackets in $\cexG$. Each face of the graph $\cexG$ is shared by 2 jackets. Summing
over the jackets, one therefore has
$\sum_{J} F_{J} = 2 F_{\cexG}$. Meanwhile, concerning number of vertices
and lines, we have  $V_{J} = V_{\cexG}$ and lines $L_{J} = L_{\cexG}$, respectively.

The next stage is to pass to topological numbers associated with the graphs. 
The Euler characteristic of a ribbon graph can be only 
defined if the graph is closed. This is the reason why we need to consider
pinched jackets $\tJ$. After closing all external half-lines of open 
jacket graphs $J$, we are in presence of closed ribbon graphs $\tJ$ for which the above topological
number is well defined.  
For the resulting jacket $\tJ$, we have 
\beq
V_{J} = V_{\tJ},  \qquad 
L_{J} = L_{\tJ} ,
\eeq 
hence $\tJ$ has the same number of vertices, 
the same number of lines as $J$,  but a different number of faces. 
The number $F_{\tJ}$ of faces of $\tJ$ finds the decomposition $F_{\tJ} = F_{\inter; \tJ} + F_{\ext; \tJ}$, where $F_{\inter; \tJ}$
is equal to  $F_{\inter; J} $ the number of internal faces of $J$
and $ F_{\ext; \tJ}$ is the number of additional closed faces
entailed by the pinching procedure. 

The Euler characteristic of $\tJ$ affords
\beq
F_{\inter;\, \tJ} + F_{\ext;\, \tJ}= 2-2g_{\tJ} - V_{J} + L_{J}. 
\label{eulertj}
\eeq 
$F_{\inter;\, \tJ}$ can be further divided into $ F_{\inter;\,   \tJ; \, \cG}$, 
the number closed faces belonging to $\cG$ and $F_{\inter;\,   \tJ; \, \cexG} $,
the number of closed faces belonging to $\cexG$ and not to $\cG$: 
\beq
F_{\inter;\, \tJ} =  F_{\inter;\,   \tJ; \, \cG} + F_{\inter;\,   \tJ; \, \cexG} .
\eeq
Summing over all  jackets, we have
\beq
\sum_{J} ( F_{\inter;\,   \tJ; \, \cG} + F_{\inter;\,   \tJ; \, \cexG} + F_{\ext; \tJ}) = 
2 F_{\inter;\cG} + 2 F_{\inter;\;  \cexG}  + \sum_{J} F_{\ext; \tJ}\;,
\label{fint}
\eeq
where, we recall that $F_{\inter;\;  \cexG}$ is the number of faces issued from $\cexG$ 
and not appearing in $\cG$ and $F_{\inter;\cG} $ is the number of closed
faces of $\cG$. 

\begin{figure}
 \centering
     \begin{minipage}[t]{.8\textwidth}
      \centering
\includegraphics[angle=0, width=12cm, height=2cm]{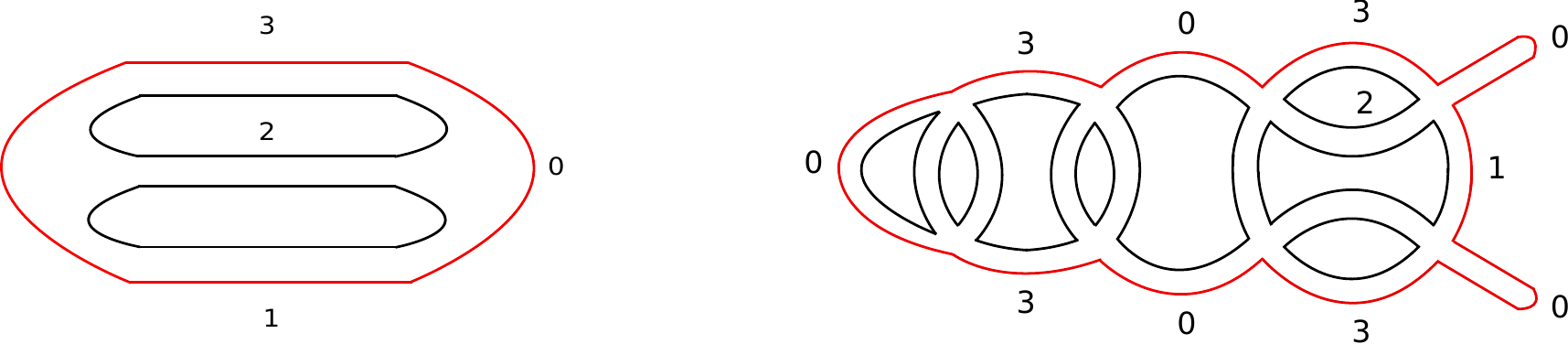}
\caption{ {\small  A face (highlighted) of $\bG$ (boundary graph of $\cG$ Fig.\ref{fig:colorext}) labeled by colors (013) and the unique pinched 
jacket $\tJ$ (0123) which possesses this face.}}
\label{fig:boundface}
\end{minipage}
\end{figure}

The quantity $F_{\inter;\;  \cexG}$  can  be computed explicitly:
each $\varphi^4$ vertex contains $4$ internal faces coming from 
the coloring (and not present in $\cG$) and 
each $\varphi^2$ type vertex contains $3$ of them. Then, one has
\beq
F_{\inter;\; \cexG}  = 4 V_4+ 3(V_2 +  \sum_{s=1}^3V'_{s;2}).
\label{fintcol}
\eeq 
Besides, using \eqref{vexlex}, we have
\beq
\sum_{ J}\left(- V_{J} + L_{J}\right)  = 3 (V_{\cexG} - \frac{1}{2}N_{\ext})
 = 
  3\left( 4V_4  +2(V_2  +\sum_{s=1}^3V'_{s;2} )  -  \frac12 N_{\ext}\right). 
\eeq
 Summing over $J$ in \eqref{eulertj}, and using the above relation,  we get
from \eqref{fint} and \eqref{fintcol}
\beq
F_{\inter;\cG}  =2 V_4  -  \frac34 N_{\ext} + 3 - \sum_{J} g_{\tJ}   -  \frac12  \sum_{J} F_{\ext; \tJ} .
\label{facint1}
\eeq
We re-express $ \sum_{J} F_{\ext; \tJ}$
 in terms of topological numbers of the boundary graph  $\bG$. 
The boundary $\bG$  is defined  from its number of vertices
$V_{\bG}$ and lines $L_{\bG} $ such that 
\beq
V_{\bG} = N_{\ext} ,\qquad
L_{\bG} = F_{\ext} .\label{vbglbg}
\eeq
The external legs of the initial graph $\cG$ have 3 strands and we have
\beq
3N_{\ext} =   2 F_{\ext}.
\label{3n2f}
\eeq
The boundary graph is a closed ribbon graph, thus, $\bG$ has a single
and closed jacket, itself. Note that a boundary graph may have several connected
components.  
We get from the Euler formula 
\beq
 2C_{\bG} - 2 g_{\bG} = V_{\bG}  -  L_{\bG}  + F_{\bG},
\eeq
where $F_{\bG}$ and $C_{\bG}$ are, respectively, the number of faces 
and of connected components of $\bG$.   It is simple to deduce
from \eqref{vbglbg} and \eqref{3n2f} that $L_{\bG} - V_{\bG} = N_{\ext}/2$ and,
from the latter and the above Euler formula, one recovers
\beq
F_{\bG}  =2(C_{\bG}-1) -2g_{\bG} + 2 + \frac{1}{2}  N_{\ext} . 
\label{fbg}
\eeq 
We remark that 
\beq
\sum_{J} F_{\ext; \,\tJ} = F_{\bG},
\label{suf}
\eeq because  each face
of the boundary graph is uniquely represented in a unique $\tJ$.
 Indeed, we recall that a face of the boundary $\bG$ can be represented by a color triple $(0ab)$; 
a face $(0ab)$ belongs to a jacket $\tJ$ if $\tJ$ is the form $(0acb)$.
A jacket being a cycle this implies that $c$ should be fixed  (See Fig.\ref{fig:boundface}).  
Then evaluating  \eqref{suf} via \eqref{fbg},  and inserting the result in \eqref{facint1}, it can be inferred

\bea
F_{\inter;\cG}  &=& 2 V_4  -  \frac34 N_{\ext} + 3 - \sum_{J} g_{\tJ}   -  \frac12 
(2(C_{\bG}-1) -2g_{\bG} + 2 + \frac{1}{2}  N_{\ext} ) \crcr
&=& 
2 V_4  -   N_{\ext} + 2 - \sum_{J} g_{\tJ} -(C_{\bG}-1)  +
g_{\bG} .
\label{facint2}
\eea
  Inserting the latter  in $ \omega_d(\cG)=-L + F_{\inter;\cG}+V'_2$, where $L=(1/2)[4 V_4  +2 (V_2  + \sum_{s=1}^3V'_{s;2} )- N_{\ext}]$,
we finally get \eqref{omeg} which achieves the proof of the theorem. 

\qed

The  quantity
\bea
-\sum_{J} g_{\tJ} + g_{\bG} - (C_{\bG}-1)
\label{gen}
\eea
 should be analyzed in detail since, mainly, the classification 
of divergent graphs relies on its behavior. The next section is devoted to this study.

\subsection{Bounds on genera}
\label{subsect:cdegan}

This section undertakes 
the study of  the quantity $-\sum_{J} g_{\tJ} + g_{\bG}$
which turns out to be the central object capturing the behavior
of the divergence degree. From that result, 
we will be able to classify the primitively divergent graphs
in a next stage.

The strategy here follows mainly the same adopted in \cite{arXiv:1111.4997}: we perform a sequence of $0k$-dipole contraction \cite{Gur3,Lins,FerriCag} of a given graph and scrutinize
the genus change under such moves. 

Let us give a flavor of the following combinatorial analysis
before starting it. A $k$-dipole contraction on a colored vacuum graph
\cite{Gur3} generalizes to the tensor case, the so-called line 
contractions along a tree for matrix ribbon graphs. Performing such a sequence erases all bubbles with melonic structure and one gets another graph
which can be called a Filk rosette \cite{Filk} of the tensor kind. The degree 
of that graph $\omega(\cG)=\sum_{J}g_{J}$ is the same as the 
degree of the initial graph since the genus
of each jacket along this sequence of contractions is preserved.

In the present case (and also in \cite{arXiv:1111.4997}), starting from a colored theory with colored graphs, 
we do not perform arbitrary $k$-dipole contractions
but, at first, $k$-dipole contractions involving all colors but 
$0$. The result is a graph with the same degree as previously claimed. 
In particular, erasing all melonic contributions yields
exactly a graph  of our starting theory. 
In a second step, we start what we call a $0k$-dipole contraction
involving now the last color $0$ (a precise definition will be given
below). Under such a contraction, the graph may or may not
change of degree. We should analyze in detail the consequence
of performing the contraction. The main points revealed
in \cite{arXiv:1111.4997} were: (1) the boundary graph coincides with
the resulting graph obtained after  any full $0k$-dipole contraction in an arbitrary sequence and removal of external structure
of a given graph; (2) under a single $0k$-dipole contraction,
the genus of a jacket never increases
and (3) the sum over all jackets of differences between genera 
of these jackets before and after contraction is always  
bounded from below by a fixed constant. 

In the present context, we actually expect the same properties
for graphs.

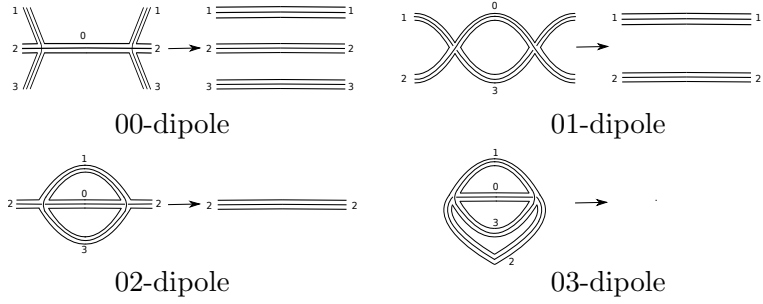
\begin{figure}
 \centering
     \begin{minipage}[t]{.8\textwidth}
      \centering
\includegraphics[angle=0, width=10cm, height=3.5cm]{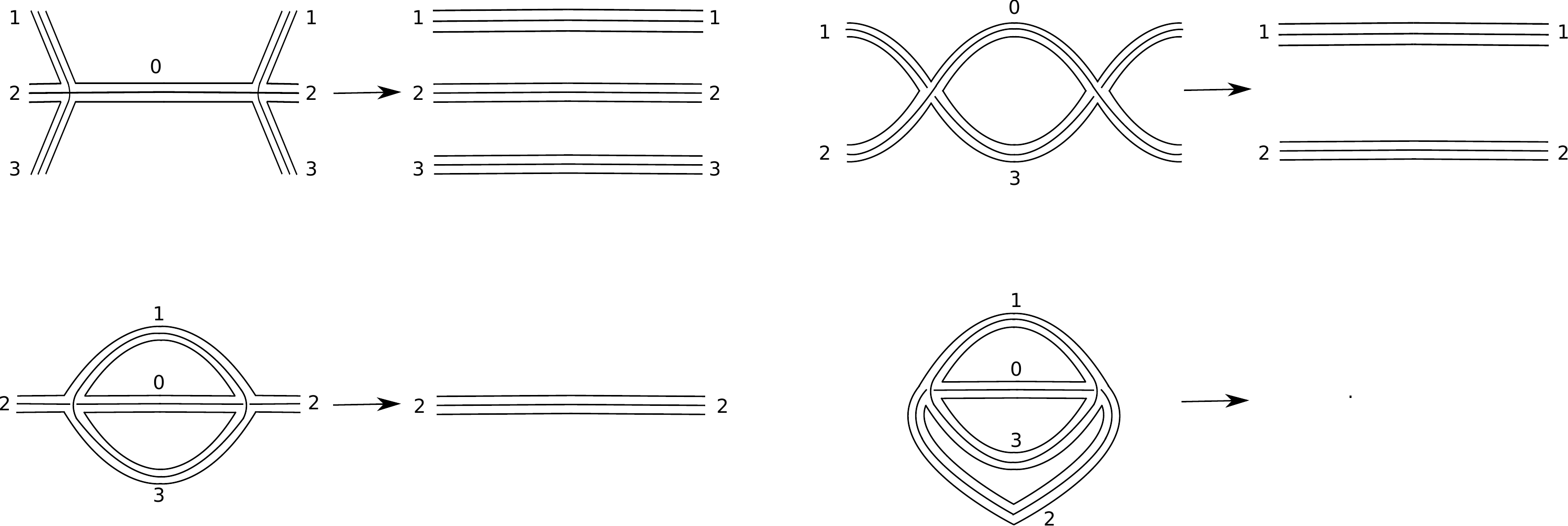}
\caption{ {\small $0k$-dipole contractions.}}
\label{fig:contract}
\end{minipage}
\put(-290,50){\small $00$-dipole}
\put(-290,-10){\small$02$-dipole}
\put(-125,50){\small $01$-dipole}
\put(-125,-10){\small $03$-dipole}
\end{figure}

\begin{definition}[$0k$-dipole and contraction]
We define a $0k$-dipole, $k=0,1,2,3$, as a maximal subgraph of 
$\cexG$ made of $k+1$ lines joining two vertices, 
one of which of color 0. Maximal means the  $0k$-dipole 
is not included in a $0(k+1)$-dipole. 

The contraction of a $0k$-dipole erases the $k+1$ lines 
of the dipole and joins the remaining $3-k$ lines on both 
sides of the dipole by respecting colors (see Fig.\ref{fig:contract}).
\end{definition}

We summarize in the following proposition a basic fact about
the boundary graph (the proof is totally similar of that 
of Lemma 3(Graph Contraction) in \cite{arXiv:1111.4997})
\begin{proposition}[Graph contraction]
\label{contract}
Performing  the maximal 
number $(4 V_4 - N_{\ext})/2$ of $0k$-dipole contractions on 
$\cexG$ in any arbitrary order and erasing the external legs 
and the remaining open faces of $\cG$ leads to the boundary graph $\bG$. 
\end{proposition}

 We restrict the analysis to the only significant
situation of graphs without any two-points vertices of the kinds $V_2$ and $V_{s;2}'$\footnote{We can first analyze graphs for which 
one performs a full and maximal contraction of all $\varphi^2$ type vertices into a single line. After, the case including such vertices
can be derived by re-inserting $\varphi^2$ vertex chains
from that point.}
and aim at studying the change in genus of a single closed jacket under a dipole contraction. We write
\beq
\cexG \to \cexG' ,\qquad J \to J' ,\qquad g_{\tJ} \to g_{\tJ'}.
\eeq
The analysis of the sum of differences between genera 
before and after the contraction will be a corollary of that result. 

Consider a colored connected  graph $ \cexG$, a fixed 
$0k$-dipole and the contracted graph  $ \cexG' $, which may or may not be connected. 
After a contraction, the numbers of vertices and lines meet the
following modifications:
\beq
V \to V' =V-2, \qquad
L \to L' = L-4 , 
\label{vv'}
\eeq
whereas the number of connected components may change from $c=1$ to    $ c' \leq 3  $.

The other ingredient entering in the definition of genus
is the number of faces. The change in faces can be
handled by another notion of pair types associated with 
dipole contractions
(see Appendix \ref{app:prooflem}). 
The key observations are summarized by the following statements:

\begin{lemma}[Decreasing genera]
\label{localjacket}
Given a pinched jacket $\tJ$, we have
\beq  g_{\tJ} - g_{\tJ'}   \geq  0 .
\label{interm1}
\eeq
\end{lemma}
\noindent{\bf Proof.} See Appendix \ref{app:prooflem}.

 \begin{lemma}[Genus bounds]
\label{genbound}
We have
\beq  \label{interm3}
g_{\tJ} \geq  g_{\bG}\;, \quad
\sum_{J} g_{\tJ} -  3 g_{\bG} \in \N .
\eeq
\end{lemma} 
 \noindent{\bf Proof.}   
A complete sequence of $0k$-dipole contractions on the graph $\cG$
yields the graph $\bG$ (see Proposition \ref{contract}).
Using Lemma \ref{localjacket}, any genus of any pinched jacket $\tJ$ decreases along that sequence of contractions. Finally,
the pinched jacket coincides with the boundary graph\footnote{ 
 This results from Proposition \ref{contract}; since
after a full sequence of contractions $\cG \to \bG$, any pinched jacket
of $\cG$ should have a projected in $\bG$. However $\bG$ is unique
so that
the conclusion is immediate.} itself $\bG$, which  proves  \eqref{interm3}. 
The factor of 3 appearing in the second statement of \eqref{interm3}, 
come from the fact that  there must exist at least three jackets $\tJ$ 
(in the sequence) generating the boundary
graph which possess a greater genus. Indeed, if $(abc)$ 
defines the boundary graph cycle (obtained after contracting the color 0,
such that $a,b,c\neq 0$), 
then three jackets 
$(0abc)$, $(a0bc)$ and $(ab0c)$ contract back on $(abc)$.
Some of the ``ancestors'' (placed upstream with respect to the 
sequence of contractions) 
of these three jackets or the three jackets themselves should possess a genus strictly greater than $g_{\bG}$. 

\qed

\subsection{Divergent graphs}
\label{subsect:digraph}

We always assume that $\bG \neq \emptyset$ hence $C_{\bG} \geq 1$. 
Furthermore $C_{\bG} \leq N_{\ext}/2$,
because each connected component must have at least 
a non zero even number of external legs.

After an analysis (the details of which are collected in 
Appendix \ref{app:conv}), the following table gives a list
of primitively divergent graphs: 

\medskip 

\begin{center}
\begin{tabular}{lcccc||cc|}
$N_{\ext}$ & $V_2 $ &  $ g_{\bG}$ & $C_{\bG}-1$ & $\sum_{\tJ}  g_{\tJ}$ & $\omega_d(\cG)$  \\
\hline\hline
4 & 0 & 0 & 0 & 0 & 0 \\
\hline
2 & 0 & 0 & 0 & 0 & 1\\
2 & 1 & 0 & 0 & 0 & 0\\
2 & 0  & 0 & 0 & 1 & 0\\
\hline\hline
\end{tabular} 
\medskip 

Table 1: List of primitively divergent graphs
\end{center}

\medskip 

We emphasize the fact that the anomalous term of 
the form $(\int \phi^2)(\int \phi^2)$  occurring 
in four dimensions \cite{arXiv:1111.4997} does not appear here.  
One could claims that in higher dimensions, more anomalous 
might be present. From the above table characterizing all dangerous contributions, 
we are in position to address the renormalization program
for these amplitudes and this is the purpose of the next
section.

\section{Renormalization}
\label{sect:renorm}

The renormalization procedure of primitively divergent 
graphs could be implement in the way of \cite{arXiv:1111.4997}
if one switches to the direct space
using techniques developed in \cite{Rivasseau:1991ub,Gurau:2005gd}. 
We equivalently perform the renormalization procedure
in the momentum space in a similar way of  \cite{Gurau:2008vd}. 

 The renormalization schemes corresponding to any 
primitively  divergent contribution given by Table 1 have to be studied. 
Nevertheless, the procedure remaining
similar at a given number of external legs, we 
will only treat the following cases: 
\begin{enumerate}
\item[(i)] $N_{\ext}=4$ yielding $\omega_d(\cG)=0$ (first line of Table 1);

\item[(ii)] $N_{\ext}=2$ yielding $\omega_d(\cG)=1$ (second line of Table 1).
\end{enumerate}
We  emphasize that even though the above cases (and for the first situation, for a given vertex configuration) will be discussed, the analysis performed in the following can be carried out
for all possible cases and affords the same conclusion.

\subsection{Renormalization of the four-point function}
\label{subsect:fourpt}

Consider a  four-point function subgraph $G^k_i$ characterized by the first line of Table 1 and equipped with four external propagators. 
The graph is such that $g_{\bG}=0$,
hence has a boundary graph of the melonic type. 
The pattern followed by external momenta 
are necessarily of the form $V_{4}$ (see Fig.\ref{fig:vertex}).
Denoting external momenta data 
by $p^{\ext}_{f}$,  $f=1,2,3,1',2',3'$, we assign
to each $p^{\ext}_{f}$  an external or boundary face (there are six of them). 
Calculations will be only made for the initial configuration 
of the $V_{4}$ vertex given by \eqref{vertex4}. It will be obvious that
for any other permutation, the derivations will lead to similar
conclusions.

Let $\cA_4(G^k_i)$ be the amplitude of the resulting
graph. Note that the external leg indices are at scale $j_l$ which must be 
strictly smaller than $i$ the scale index of the quasi-local 
subgraph $G_i^k$. 

Using the face factorization,
$\cA_4(G^k_i)$ can be written  
(alleviating notations, henceforth $p_s$ refers directly 
to the absolute value $|p_s|$ and the reference graph $G^k_i$ will be dropped) as
\bea
\cA_4[\{ p^\ext_{f}\}  ] = 
\sum_{ p_{f} } 
\int [\prod_{\ell} d\alpha_\ell e^{-\alpha_\ell m} ]
\Big\{ 
\Big[   \prod_{ f \in F_{\ext} } 
e^{ - (\sum_{ \ell \in f } \alpha_\ell  )  a_fp^{\ext}_f  } 
\Big]
\Big[ \prod_{ f \in F_{\inter} } 
e^{- (\sum_{\ell \in f} \alpha_\ell  )a_fp_f } 
\Big] 
\Big\}.
\label{ampl4p}
\eea 
 each $\alpha_\ell \in [M^{-d_\ell},M^{-d_\ell +1}]$,  $\ell$
runs over all lines of the graph. Referring to an external 
line, we will use instead $l$.  Hence, 
$d_\ell$ can be either an external index $j_l$ or an internal 
one $i_\ell$. In particular, for all $f\in F_{\inter}$ in the second product, $\alpha_\ell $
is at scale $i_\ell \gg j_l$.

Next, we single out for each independent external face the exponent 
$\alpha_{l} p^\ext_f$, where $\alpha_{l} \in [M^{-j_l},M^{-j_l +1}]$, 
such that, we can write a given external face amplitude as
\bea
e^{ - (\sum_{ \ell \in f } \alpha_\ell  ) a_f  p^{\ext}_f  } 
 = e^{-(\alpha_{l} + \alpha_{l'} )a_f p^{\ext}_f } 
e^{ - (\sum_{ \ell \in f / \ell \neq l } \alpha_\ell  ) a_f  p^{\ext}_f  } .
\eea
Noting that all $\alpha_{\ell\neq l}$, in the sum, are now at scale $i_\ell \gg j_l$, 
and $p_f^\ext \sim M^{j_l}$, we perform a Taylor expansion
for each external face as
\beq
e^{ - (\sum_{ \ell \in f } \alpha_\ell  )a_f p^{\ext}_f  } 
 = e^{-(\alpha_{l} + \alpha_{l'} ) a_f p^{\ext}_f } 
\Big[ 1  - (\sum_{ \ell \in f / \ell \neq l } \alpha_\ell  )  a_f p^{\ext}_f 
\int_0^1 dt 
e^{ -t\, (\sum_{ \ell \in f / \ell \neq l } \alpha_\ell  ) a_f  p^{\ext}_f  }  \Big].
\label{taylor4}
\eeq
Rewriting \eqref{taylor4} 
as $e^{ - (\sum_{ \ell \in f } \alpha_\ell  ) a_f  p^{\ext}_f  } 
 = e^{-(\alpha_{l} + \alpha_{l'} )a_f  p^{\ext}_f } 
[ 1  - R^\ext_f],$ where $R^\ext_f$ is the remainder of the
Taylor expansion, we substitute the latter expression 
in the initial amplitude $\cA_4[\{ p^\ext_{f}\}  ] $. 
In loosely notations, the result is
\bea
&&
\cA_4[\{ p^\ext_{f}\}  ] =\sum_{ p_{f} } 
\int [\prod_{\ell} d\alpha_\ell e^{-\alpha_\ell m} ]
\Big\{ 
 \label{ampl4prr}\\
&&
 \Big[\prod_{ f \in F_{\ext} } e^{-(\alpha_{l} + \alpha_{l'} ) a_f p^{\ext}_f }
\Big] \Big[ 1-  \sum_{f \in F_{\ext}} R^\ext_f + \sum_{f,f' \in F_{\ext}} R^\ext_f R^\ext_{f'}
- \ldots 
\Big] 
\Big[ \prod_{ f \in F_{\inter} } 
e^{- (\sum_{\ell \in f} \alpha_\ell  )a_f p_f } 
\Big] 
\Big\}.
\nonumber
\eea 
Collecting the leading order contribution, one has
\bea
&&
\cA_4[\{ p^\ext_{f}\} ;0 ] = 
\sum_{ p_{f} } 
\int[\prod_{\ell} d\alpha_\ell e^{-\alpha_\ell m} ]
\Big\{ 
\Big[   \prod_{ f \in F_{\ext} } 
e^{-(\alpha_{l} + \alpha_{l'} ) a_f p^{\ext}_f } 
\Big]
\Big[ \prod_{ f \in F_{\inter} } 
e^{- (\sum_{\ell \in f} \alpha_\ell  )a_f p_f } 
\Big] 
\Big\} 
\label{vertexren}\\
&& = 
 \Big\{
\int [\prod_{l} d\alpha_l e^{-\alpha_l m} ]
 \prod_{ f \in F_{\ext} } 
e^{-(\alpha_{l} + \alpha_{l'} ) a_f p^{\ext}_f } 
 \Big\}  \Big\{ 
 \sum_{ p_{f} } 
\int [\prod_{\ell \neq l} d\alpha_\ell e^{-\alpha_\ell m}]
\prod_{ f \in F_{\inter} } 
e^{- (\sum_{\ell \in f} \alpha_\ell  )a_f p_f } 
\Big\} .
\nonumber
\label{ampl4p0}
\eea 
The first factor corresponds to a vertex hooked
to four external propagators since it can be recast
in a mere form
\bea
&&
\int [\prod_{l} d\alpha_l e^{-\alpha_l m} ]
\Big[   \prod_{ f \in F_{\ext} } 
e^{-(\alpha_{l} + \alpha_{l'} ) a_f p^{\ext}_f } 
\Big] \crcr
&&=  \int [\prod_{l} d\alpha_l]\Big\{
 e^{-\alpha_{l_1} (a_1p_1 + a_2p_2 + a_3p_3 +m)} 
 e^{-\alpha_{l_2} (a_1p'_1 + a_2p_2 + a_3p_3+m)}\\
&&
 e^{-\alpha_{l_3} (a_1p_1' + a_2p_2' +a_3 p_3'+m)}
e^{-\alpha_{l_4} (a_1p_1 + a_2p_2' + a_3p_3'+m)}\Big\}.
\nonumber
\eea
The second factor coincides with the logarithmically divergent
internal contribution given by the power-counting theorem. 
Hence $\cA_4[\{ p^\ext_{f}\} ;0 ]$ determines a log-divergent
counterterm participating in the vertex renormalization.

The remaining terms in \eqref{ampl4prr} should be analyzed. 
Proving that the first subleading order contribution improves the power-counting 
will be a sufficient condition to ensure the convergence of the subsequent product terms. The first subleading term is of the form
\bea
&&
R_4 =
\sum_{ p_{f} } 
\int [\prod_{\ell} d\alpha_\ell e^{-\alpha_\ell m} ]
\Big\{ \Big[\prod_{ f \in F_{\ext} } e^{-(\alpha_{l} + \alpha_{l'} ) a_fp^{\ext}_f }
\Big]\\
&&
 \times\Big[  -\sum_{f \in F_\ext} (\sum_{ \ell \in f / \ell \neq l } \alpha_\ell  )  a_fp^{\ext}_f 
\int_0^1 dt 
e^{ -t\, (\sum_{ \ell \in f / \ell \neq l } \alpha_\ell  ) a_f p^{\ext}_f  }  \Big] \Big[ \prod_{ f \in F_{\inter} } 
e^{- (\sum_{\ell \in f} \alpha_\ell  )a_fp_f } 
\Big] 
\Big\} \nonumber
\eea 
and can be bounded by
\beq
|R_4| \leq K M^{-(i(G^k_i) - e(G^k_i))}
\sum_{p_f} \int [\prod_{\ell} d\alpha_\ell e^{-\alpha_\ell m} ]
\Big[ \prod_{ f \in F_{\inter} } 
e^{- (\sum_{\ell \in f} \alpha_\ell  )a_fp_f } 
\Big] ,
\eeq
where $K$ is some constant  (which could depend
on the number of internal lines of the graph and number of
 external faces).
Note that the integral over $t$ yields just a factor of $O(1)$. 
Also remark that, we used 
optimal bounds such that $p^\ext_f \leq  M^{e(G^k_i)}$, 
and, for any internal decay, one assumes that $\alpha_\ell \leq M^{-i(G^k_i)}$,
recalling that $e(G^k_i) = \sup_{l} j_l$
and $i(G^k_i) = \inf_{\ell \in G^k_i} i_\ell$.
The factor $M^{-(i(G^k_i) - e(G^k_i))}$ guarantees that the power-counting  $M^{\omega_d(G^k_i)=0}$ (which corresponds to the last sum, 
up to some constant) is improved
and will bring the sufficient decay to ensure the sum of momentum attributions in the standard way of \cite{Rivasseau:1991ub}.

\subsection{Renormalization of the two-point function}
\label{subsect:2pt}

We consider a graph $G^k_i$ with two external 
legs with topology as dictated by one the corresponding 
rows of Table 1. Note that the boundary data $p^\ext_{f}$, $f=1,2,3,$
now follow the pattern of a simple mass vertex. 
Only, the case of a two-point graph with a linear divergence will be treated since the subtraction of log-divergent cases 
can be recovered easily from the same analysis by restricting the 
Taylor expansion at less order. 
 
Let $\cA_2[\{p^\ext_f\}]$ be 
the full amplitude associated with the graph $G^k_i$
with external propagators.  
Having defined all tools in the previous analysis of Subsection \ref{subsect:fourpt}, 
we perform the following expansion of any boundary 
face amplitude occurring in $\cA_2[\{p^\ext_f\}]$ as
\bea
&&
e^{ - (\sum_{ \ell \in f } \alpha_\ell  )  a_fp^{\ext}_f  } 
 = e^{-(\alpha_{l} + \alpha_{l'} ) a_fp^{\ext}_f } \label{taylor2}\\
&&
\times
\Big[ 1  - (\sum_{ \ell \in f / \ell \neq l } \alpha_\ell  )  a_fp^{\ext}_f 
 +
 \big[  - (\sum_{ \ell \in f / \ell \neq l } \alpha_\ell  )  a_fp^{\ext}_f \big]^2
\int_0^1 dt (1-t)
e^{ -t\, (\sum_{ \ell \in f / \ell \neq l } \alpha_\ell  ) a_f p^{\ext}_f  }  \Big].
\nonumber
\eea
The same \eqref{taylor2} can rewritten as $e^{ - (\sum_{ \ell \in f } \alpha_\ell  ) a_fp^{\ext}_f  } 
=e^{-(\alpha_{l} + \alpha_{l'} )a_f p^{\ext}_f } [1+ S^\ext_f + R_f^\ext ].$
Substituting this expression in the initial amplitude yields
\bea
&&
\cA_2[\{p^\ext_f\}]  =  
\sum_{ p_{f} } 
\int [\prod_{\ell} d\alpha_\ell e^{-\alpha_\ell m} ]
\Big\{ \crcr
&&
\Big[ \prod_{ f \in F_{\ext} } e^{-(\alpha_{l} + \alpha_{l'} )a_f p^{\ext}_f }  \Big] \Big[
1+ \sum_{f\in F_{\ext}  } S^\ext_f + \sum_{f\in F_{\ext} } 
R^\ext_f  \crcr
&&+ \sum_{f,f'\in F_{\ext} } (S^\ext_f + R^\ext_{f})(S^\ext_{f'} +R^\ext_{f'}) +  \dots 
\Big]
\Big[ \prod_{ f \in F_{\inter} } 
e^{- (\sum_{\ell \in f} \alpha_\ell  )a_fp_f } 
\Big] 
\Big\}.
\label{ampl2p}
\eea
As expected, the leading order contribution $\cA_2[\{ p^\ext_{f}\} ;0 ] $ 
is mainly of the factorized form \eqref{vertexren} with, in the present instance, the first factor appearing as
\bea
&&
\int [\prod_{l} d\alpha_l e^{-\alpha_l m} ]
\Big[   \prod_{ f \in F_{\ext} } 
e^{-(\alpha_{l} + \alpha_{l'} ) a_f p^{\ext}_f } 
\Big] \crcr
&&=
\int [\prod_{l} d\alpha_l]
 e^{-\alpha_{l_1} (a_1p_1 + a_2p_2 +a_3 p_3+m)} 
 e^{-\alpha_{l_2} (a_1p_1 + a_2p_2 +a_3 p_3+m)}
\eea
and the second factor bringing a linear divergence. Clearly, this 
term belongs to a mass renormalization. 

The remaining task is to prove that higher order terms
improve the power-counting. Let us focus on the following
\bea
&&
\cA_2'[\{p^\ext_f\};0]  =  
\sum_{ p_{f} } 
\int [\prod_{\ell} d\alpha_\ell e^{-\alpha_\ell m} ]
\Big\{ \crcr
&&
\Big[ \prod_{ f \in F_{\ext} } e^{-(\alpha_{l} + \alpha_{l'} )a_f p^{\ext}_f }  \Big] \Big[ -\sum_{f \in F_\ext}   (\sum_{ \ell \in f / \ell \neq l } \alpha_\ell  )a_fp^{\ext}_f  \Big] 
\Big[ \prod_{ f \in F_{\inter} } 
e^{- (\sum_{\ell \in f} \alpha_\ell  )a_fp_f } 
\Big] 
\Big\}
\label{ampl2p0}
\eea
which can be seen as a wave-function renormalization. 
Indeed, we can recompose it as 
\bea
&&
\cA_2'[\{p^\ext_f\};0]  =  -\sum_{s=1}^3 \Bigg\{ 
 \crcr
&&
 \Big[ \int [\prod_{l} d\alpha_l  ]
e^{-\alpha_{l_1} (a_1p_1 + a_2p_2 +a_3 p_3+m)} 
 e^{-\alpha_{l_2} (a_1p_1 + a_2p_2 +a_3 p_3+m)}a_sp^{\ext}_s
 \Big] 
\crcr
&&\Big[\sum_{ p_{f} } \int [\prod_{\ell \neq l} d\alpha_\ell e^{-\alpha_\ell m} ]\Big(  \sum_{  \ell \in f_s / \ell \neq l } \alpha_\ell   \Big)
 \prod_{ f \in F_{\inter} } 
e^{- (\sum_{\ell \in f} \alpha_\ell  )a_fp_f } 
\Big] \Bigg\},
\label{ampl2p00}
\eea
where the first factor is nothing but a wave function type contribution
(and renormalizes $a_s$) and the second factor is log-divergent. The latter statement is justified by the fact that
the factors  $\sum_{ \ell \in f_s / \ell \neq l } \alpha_\ell $
brought by the derivatives are of order $M^{-i_\ell}$ which
annihilates the internal divergence degree which was linear.

As another consequence of the decay of $\sum_{ \ell \in f / \ell \neq l } \alpha_\ell $, any monomial in $S_f$ of order
higher than 2  in \eqref{ampl2p} is simply convergent. 

Next, the first order remainder term is  of the form 
\bea
&&
R_2 = 
\sum_{ p_{f} } 
\int [\prod_{\ell} d\alpha_\ell e^{-\alpha_\ell m} ]
\Big\{  \Big[\prod_{ f \in F_{\ext} } e^{-(\alpha_{l} + \alpha_{l'} )a_f p^{\ext}_f }  \Big] \label{ampl2rem}\\
&&
\sum_{f\in F_{\ext}}\Big[\big[ (\sum_{ \ell \in f / \ell \neq l } \alpha_\ell  )  a_fp^{\ext}_f \big]^2
\int_0^1 dt (1-t)
e^{ -t\, (\sum_{ \ell \in f / \ell \neq l } \alpha_\ell  ) a_f p^{\ext}_f  } \Big] 
\Big[ \prod_{ f \in F_{\inter} } 
e^{- (\sum_{\ell \in f} \alpha_\ell  )a_fp_f } 
\Big] 
\Big\}.
\nonumber
\eea
Using the fact that $p^\ext_f \leq  M^{-e(G^k_i)}$ whereas
$\alpha_\ell \sim M^{-i_\ell}$, with $i_\ell \gg e(G^k_i)$, for $\ell \neq l$, we find the following optimal bound for the above term as
\bea
|R_2| \leq K M^{-2(i(G^k_i) - e(G^k_i))}
\sum_{ p_{f} } 
\int [\prod_{\ell \neq l} d\alpha_\ell e^{-\alpha_\ell m} ]
\Big[ \prod_{ f \in F_{\inter} } 
e^{- (\sum_{\ell \in f} \alpha_\ell  )a_fp_f } 
\Big] ,
\eea
where the integral over $t$ yields again a factor of $O(1)$.
We recognize, in the last sum, the linear divergence associated
with the two-point graph at $t=0$. Hence, this term is 
convergent and will ensure the summability of scale attributions. 

All the rest of terms are manifestly convergent
from the fact that they involve only product of convergent
terms.
Finally, the summability of the scale attributions 
can be performed in the way introduced in \cite{Rivasseau:1991ub}.

\subsection{About enlarging and reducing the space of couplings}
\label{subsect:collaps}

It is worthwhile to discuss how the (perturbative) renormalizability of the initial model can be extended to different models with various degrees of anisotropy between colors. Different classes of models
are obtained by just putting a different coupling for each term or conversely identifying the different couplings. 

Renormalizability is triggered by the coherent combination of three ingredients: a multi-scale analysis, a locality principle and a power-counting theorem. The issue raised above can be rephrased as ``is a given class of theory with particular anisotropies stable or not under
the RG flow?''. 

Consider the renormalizable model defined by the action written in terms of \eqref{eq:action0} and \eqref{vertex4}, which
we call in the following discussion (I). The model has many wave function couplings, namely $a_s$, and one single interaction coupling, $\lambda$
(the mass is always kept fixed in the next developments). 

In addition to the model (I), let us discuss three other basic classes
of models:

(1) The most general ``anisotropic'' model which has different couplings for each 
interaction and different wave function couplings;
hence, in this model, the main parameters are
($\lambda_{\epsilon=1,2,3}, a_{\epsilon=1,2,3}$);

(2) The  model with``anisotropic interactions" which has
different interaction couplings but 
a single wave function coupling ($\lambda_{\epsilon=1,2,3}, a_{\epsilon=1,2,3}=a$); 

(3) The ``isotropic'' model which has a single coupling for interactions 
and a single coupling for kinetic terms ($\lambda_{\epsilon=1,2,3}=\lambda, a_{\epsilon=1,2,3}=a$).

Note that all these theories are continuously connected
by parameter deformations. However, the interesting question is whether these classes are stable under the renormalization group flow. 

All models (1)-(3) share a common feature with the model (I) studied so far: 
their multi-scale analysis and power-counting theorem are identical
 since those are written without the explicit knowledge
of the coupling constants. All what is required to perform 
a power-counting is the type of propagator decay, the vertices and lines combinatorial properties and gluing rules and these
are exactly the same for these models. One can call these theories ``power-counting renormalizable'' to this respect. 

However, as previously emphasized, the renormalizability is also about a
locality principle that ought to be satisfied by the model. This feature can be also called  ``replicability'' or ``surviving
property'' along scales of the model.  
The locality principle in a just renormalizable model with marginal terms 
with $n$-valent vertices ensures, roughly, that any log-divergent $n$-point function with external data of the same form of a given interaction  should renormalize the coupling of this interaction. This principle holds immediately either for a unique coupling or different ones for marginal terms. Indeed, one has just to decide, in any situation, to affect to the renormalized coupling constant of a given vertex all divergent terms with akin external data. Even for a unique coupling shared by several interactions, this can be explicitly done. Of course, from the point of view of the RG flow, the RG equations can be very involved (the following analysis performed at one-loop does not displays this feature but it certainly occurs are further loops) and each renormalized coupling is generally dependent of many
(if not all) couplings. The key point is that they can be, at least, written
in an unambiguous way, for one or many coupling constants 
for the interaction. 

The actual and only issue during the renormalization occurs potentially with the locality principle for relevant terms.
In the present story, relevant terms are  two-point functions\footnote{In several other contexts, they are always the most diverging ones.} 
given by the second line of Table 1. 
Note that the second and third lines consist in marginal 
two-point functions and so can be simply handled by the mass
renormalization.  

Consider model (I) as starting
point. Performing a Taylor expansion of two-point functions around their local part (in the way already introduced in Subsection \ref{subsect:2pt}) yields
two important corrections: a 0-th order linearly divergent 
contribution which again contributes to the mass and a first order log-divergent
contribution involving $a_s p^\ext_s$. 
For each two-point function, one finds that the second kind of correction
has two features: 
(A) it can be identified with a term present in the Lagrangian and (B)
it should contribute to the wave function renormalization $Z_s$
of the model. 

Note that, for theories (I) and 
(1) having many coupling constants $a_{\epsilon=1,2,3}$, there is no
ambiguity to renormalize independently each of these couplings
by the above log-contribution of any two-point function and, so, to
define $Z_s$. 
This is totally independent of the presence of several coupling $\lambda_{\epsilon=1,2,3}$. 
It becomes immediate that (I) and (1)  satisfy a locality principle
for two-point functions and therefore are renormalizable. 

Let us focus on model (3) assuming that $\lambda_{\epsilon}=\lambda$, and $a_\epsilon=a$. Condition (A) above 
is fulfilled since \eqref{ampl2p00} can be reached in any case. 
However, due to the fact that we have a single wave function coupling, 
Condition (B) changes drastically: 
It should be replaced by (B'): the local log-divergent contribution of
the two-point function should define a unique wave function renormalization $Z$ independent of the strand index $s$. 
One notes from \eqref{ampl2p00}, that the log-divergent 
term actually involves the strand index $s$ (see the last line where 
$\sum_{\ell \in f_s/\ell \neq l}\alpha_\ell$ is integrated).
This issue can be overcome 
by a permutation of the vertices present in the theory.  
The latter statement is understood in the sense that given a graph which is not symmetric in the strands, there always exists a set of graphs (with the same number of vertices, lines and the same degree of divergence) identical by color permutation to the former.
This simply holds by permutation  the vertex indices $V_{4;1}\to V_{4;2} \to V_{4;3}$. 
Hence, if the next order correction of a two-point graph 
has an amplitude $\cA_2'$ \eqref{ampl2p00} depending on $s=1$, the same correction
has two partners of the same form for $s=2,3$. 
Adding these restores the symmetry in $s$ and entails that 
 the self-energy $\Sigma[p^\ext_1,p^\ext_2,p^\ext_3]$ (the sum of all amputated 1PI two-point functions at external momenta $p^\ext_s$) should be a symmetric function in all strands. 
The wave function renormalization 
\bea
Z &=& 1- \frac{1}{a}\partial_{p^\ext_1} \Sigma[p^\ext_1,p^\ext_2,p^\ext_3]|_{p^\ext_{s'}=0} =  1- \frac{1}{a}\partial_{p^\ext_2} \Sigma[p^\ext_1,p^\ext_2,p^\ext_3]|_{p^\ext_{s'}=0}\crcr
&= &1- \frac{1}{a}\partial_{p^\ext_3} \Sigma[p^\ext_1,p^\ext_2,p^\ext_3]
|_{p^\ext_{s'}=0}
\eea
should not depend on $s$. By expanding $\Sigma[b^\ext_1,b^\ext_2,b^\ext_3]$ order by order and 
even though the latter is symmetric, the fact that $Z$ 
keeps a fixed value for all derivatives $\partial_{b^\ext_s}$
cannot be met unless all coupling coincide: $\lambda_{\epsilon} = \lambda.$ This condition clearly holds for the model (3) and therefore the latter is renormalizable. In contrast, the model (2) having distinct $\lambda_\epsilon$ is not stable under RG flow: starting from
a bare action in the class (2), one ends up with an effective action
in the general class (1).

Let us discuss, finally, the following peculiar partly anisotropic theory

(4) Consider the  model with two  interaction couplings  
and two different wave function couplings such that
$\lambda_{1}\neq \lambda_{2}=\lambda_3,\; a_{1}\neq a_2 = a_3$;
note that we could have chosen any symmetric situation.

The model (4) is interesting because, we do not choose
the easy case to let all $a_s$ to be different (which will obviously
lead to renormalizability along with theories (I) and (1)), 
but we choose instead two wave function couplings
$a_2 = a_3 \neq a_1$. This model is renormalizable 
because it lies exactly in between the two different 
situations so far carried out. 

Merging the sector $a_2=a_3$ and following step by step the above analysis, it is direct to find the necessary condition 
 $\lambda_2=\lambda_3$ for having a unique wave function 
renormalization $Z_2=Z_3$. We conclude that the model (4) is renormalizable.

The model (4) is interesting by itself for the remaining analysis 
on $\beta$ function.  Indeed, it is  related
to two renormalizable models which are, furthermore, asymptotically free
(see next section): 

(4') $\lambda_{1}\neq \lambda_{2}=\lambda_3=0,\; a_{1}\neq a_2 = a_3$;

(4'') $0=\lambda_{1}\neq \lambda_{2}=\lambda_3,\; a_{1}\neq a_2 = a_3$.

It is not excluded that there exist other renormalizable theories
in this framework. The above analysis  illustrates the richness of tensor field
theories.

\section{one-loop $\gamma$-, $\beta$-functions
and RG flows }
\label{sect:beta}

This section starts the second part of our analysis which
aims at calculating the so-called $\gamma$- and $\beta$-functions governing 
coupling constant RG flows of the model. We will restrict 
the study at one-loop since it is well-known that important
properties (like asymptotic freedom for instance) of the model can be inferred even at this approximation.  

The $U(1)^3$ model described so far has a unique coupling
constant associated with all vertices $V_{4}$. It is interesting to relax
that condition and study the same model in full generality 
by assigning to each vertex a different coupling constant.   
According to the discussion of Subsection \ref{subsect:collaps}, 
the model obtained in this way is again renormalizable.
The RG equations for the renormalizable 
model (3) (from Subsection \ref{subsect:collaps}) 
can be inferred from this extended model.  
Meanwhile, RG equations for models (4') and (4') 
have to be computed with  slightly modified method. 
Moreover, we introduce a symmetry factor due to 
the internal symmetry of the vertices: each coupling constants
will be of the form $-\lambda_\epsilon/2$, $\epsilon=1,2,3$.

In the following, the basic quantities (the self-energy and
$\Gamma^4$-function) are first computed in full generality 
(i.e. within the framework of model (1))
and, then, we will particularize these on three reduced theories given by
\begin{enumerate}
\item[(i)] $[a_{\epsilon} = a,\;\; \lambda_\epsilon = \lambda]$
\item[(ii)] $[a_{2} = a_3 \neq a_1,\;\; \lambda_{2,3} = 0,\;\; \lambda_1 \neq 0]$
\item[(iii)] $[a_{2} = a_3 \neq a_1,\;\;  \lambda_{2}=\lambda_3 \neq 0,\;\;  \lambda_{1} = 0]$
\end{enumerate}
The  instance (i) corresponds to an interesting model 
with  a unique coupling constant for all interactions \eqref{vertex4} and a 
unique wave function coupling. It is simply the most natural
model which can be viewed as the rank-3 analogue of the model
investigated in \cite{arXiv:1111.4997}. On the other hand,
the second and third correspond to some more drastic truncation
that one could perform. The latter (ii) and (iii) share some but not all features 
of the former hence are different.

\subsection{$\gamma$- and $\beta$-functions}
\label{subsect:beta}

In full generality (working in the model (1) Subsection \ref{subsect:collaps}),
considering three wave function renormalizations
 $Z_{\epsilon=1,2,3}$, each of one with respect to each of the 
strands,
the field strength can be modified as follows:
\beq
\varphi \longrightarrow \left(Z_1Z_2Z_3\right)^{\frac16} \varphi,
\eeq
so that, after renormalization, the  wave function couplings 
satisfy  the equations
\beq
a_{\epsilon}^{\ren} = a_{\epsilon} \left( \frac{Z_\epsilon^2}{Z_{\check \epsilon} Z_{ \check{\check\epsilon} } } \right)^{\frac13},\; \qquad \epsilon = 1,2,3, \qquad 
\check \epsilon \neq \epsilon, \qquad 
\check \epsilon \neq \check{\check \epsilon} 
\neq  \epsilon .
\label{wcoup}
\eeq
For a rank three tensor model as the one
we are presently dealing with, we define 
\beq
Z_{\epsilon} = 1 -  \frac{1}{a_{\epsilon}}   \partial_{b_{\epsilon}} 
\Sigma \Big|_{b_{1,2,3} = 0},
\label{wfr}
\eeq
where the self-energy  $\Sigma(b_1,b_2,b_3)$ is 
the sum of the amputated 
one particle irreducible (1PI) amplitudes of the two-point correlation function truncated at one-loop that we denote
\begin{eqnarray}
 \Sigma (b_1,b_2, b_3) = \langle \, \bar\varphi_{b_1b_2 b_3}\,\varphi_{b_1b_2 b_3} \, \rangle^{t}_{1PI}.
\label{topt}
\end{eqnarray}
Our initial goal is to compute at one-loop the dynamics of the effective
couplings $a_\epsilon$ governed by the $\gamma_\epsilon$-functions 
encoded in \eqref{wcoup}. In a second stage, we will 
compute, still at one-loop, the mass $\beta_m$-function given by the expression  
\beq
\label{mreninit}
m^\ren = \frac{m - \Sigma(0,0,0)}{\left(Z_1Z_2Z_3\right)^{\frac13} }.
\eeq 
In the final step,
we will study the dynamics of constant couplings $\lambda_{\epsilon}$, 
governed by the $\beta_\epsilon$-functions ciphered by the
following equations
\begin{eqnarray}
\label{gamma4}
\lambda_{\epsilon}^{\ren} = -\frac{\Gamma_{4, \epsilon}(0,0,0,0,0,0)}{
(Z_1 Z_2 Z_3)^{\frac{2}{3}}},\; \qquad \epsilon = 1,2,3.
\end{eqnarray}
where $\Gamma_{4, \epsilon}(b_1,b_2, b_3,b_1',b_2', b_3')$
is the amputated 1PI four-point function. 

Let us recall that the amputated 1PI four-point functions 
occurring in (\ref{gamma4}) read 
\begin{eqnarray}
&&\Gamma_{4,1 }(b_1,b_2, b_3,b_1',b_2', b_3') = 
\langle \, \varphi_{b_1b_2 b_3}\,\bar\varphi_{b_1'b_2 b_3}\, \varphi_{b_1'b_2' b_3'}\,\bar\varphi_{b_1b_2' b_3'}  \, \rangle^{t}_{1PI},\cr\cr
&&\Gamma_{4, 2}(b_1,b_2, b_3,b_1',b_2', b_3') = 
\langle \, \varphi_{b_1b_2 b_3}\,\bar\varphi_{b_1b_2' b_3} \,\varphi_{b_1'b_2' b_3'}\,\bar\varphi_{b_1'b_2 b_3'}  \, \rangle^{t}_{1PI},\cr\cr
&&\Gamma_{4,3 }(b_1,b_2, b_3,b_1',b_2', b_3') = 
\langle \, \varphi_{b_1b_2 b_3}\,\bar\varphi_{b_1b_2 b_3'} \,\varphi_{b_1'b_2' b_3'}\,\bar\varphi_{b_1'b_2' b_3}  \, \rangle^{t}_{1PI},
\label{fopt}
\end{eqnarray}
where external indices, even though repeated, are not summed
and follow the pattern of the vertices of the model. This is
justified by the renormalization prescription.

Considering the reduced cases (ii) and (iii), there
are potentially two wave function renormalizations $Z_{\epsilon=1,2}$.
Nevertheless, the cancellation of one or many couplings 
may have drastic consequences on the way that the $\beta$-function
equations have to be written. We will deal with these 
after computing the self-energy as well as the $\Gamma^4$-function in full generality and then putting to zero some of  the contributions.

For $\epsilon,\check \epsilon, \check{\check \epsilon}  = 1,2,3,$
and pairwise distinct, we introduce the following formal sums:
\bea
\mathcal{S}_{\epsilon}(b)&:=&\sum_{p_{1},p_{2}\in \Z} 1/(a_\epsilon|b|+a_{\check\epsilon}|p_{1}|+a_{\check{\check\epsilon}}|p_{2}|+m) , 
\label{sepbe} \\
\mathcal{S}'_\epsilon( b, b')&:=&\sum_{p\in \Z} 1/(a_{\check \epsilon}|b|+a_{\check{\check \epsilon}}|b'|+a_{\epsilon}|p|+m), 
\label{seppbe}\\
S_\epsilon &:=&  
\sum_{p_{1},p_{2}\in \Z} 1/(a_{\check\epsilon}|p_{1}|+a_{\check{\check\epsilon}}|p_{2}|+m)^2 ,
\label{sep}
\\
S_1'
&=& \sum_{p_1,p_2} 1/(a_2 (|p_1| + |p_2|) + m)^2 , \;\quad 
\; 
S_2' = \sum_{p_1,p_2} 1/(a_1 |p_1| + a_2 |p_2| + m)^2,
\label{sprim}
\\
\underline{\mathcal S}_{\epsilon}(b,b') &:=& \sum_{p_1,p_2} 1/[(a_\epsilon |b|+a_{\check\epsilon}|p_1|+a_{\check{\check\epsilon}}|p_2|+m)(a_\epsilon|b'|+a_{\check\epsilon}|p_1|+a_{\check{\check\epsilon}}|p_2|+m)]
\label{undes}
\eea
Note that $\mathcal{S}_\epsilon$ is linearly divergent
whereas $\mathcal{S}'_{\epsilon}$, $S_\epsilon$, $S'_{1,2}$ and 
$\underline{\mathcal S}_{\epsilon}$ are logarithmically divergent. 
Another important fact to notice is  
$\underline{\mathcal S}_\epsilon(0,0) =  S_{\epsilon}$.

\begin{figure}
 \centering
    \begin{minipage}[t]{.7\textwidth}
     \centering
 \includegraphics[angle=0, width=6.5cm, height=1.5cm]{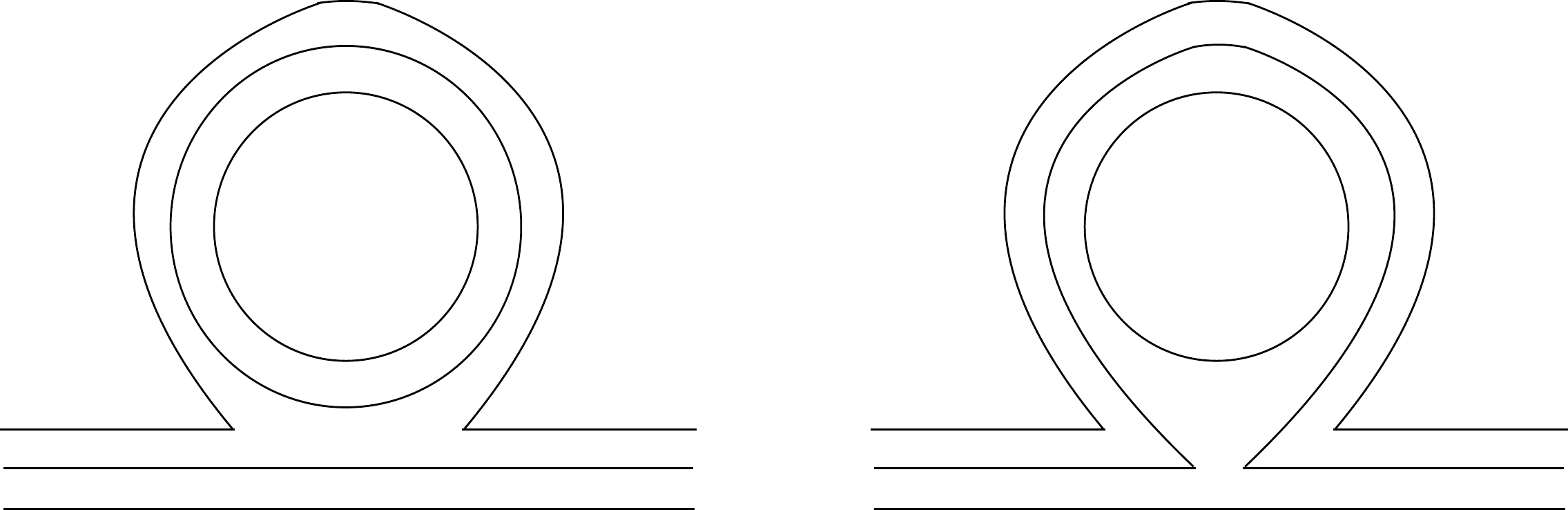}
 \vspace{0.1cm}
 \caption{ {\small Tadpoles for the first type
of vertex. }}
\label{fig:tadpoles}
 \end{minipage}
 \put(-222,-12){$T_1$}
 \put(-118,-12){$T_1'$}
 \end{figure}

Let us prove the following proposition
\begin{lemma}
\label{self}
At one-loop, the self-energy and wave function renormalizations 
are given by
\bea
\Sigma(b_1,b_2,b_3) &=& - \sum_{\epsilon} \lambda_{\epsilon} 
\mathcal{S}_{\epsilon} (b_\epsilon)  -\lambda_1\mathcal{S}'_1( b_{2}, b_{3} )   
-\lambda_2\mathcal{S}'_2( b_{1}, b_{3} )   
-\lambda_3\mathcal{S}'_3( b_{1}, b_{2} )
+ O(\lambda^2) , 
\cr\cr
Z_\epsilon &=&  1-  \lambda_\epsilon\,  S_\epsilon 
 +O(\lambda^2) ,
 \qquad \epsilon=1,2,3,
\eea
where $O(\lambda^2)$ is a  O-function of any quadratic products of any coupling constants
$\lambda_{1,2,3}$.   
\end{lemma}
\noindent{\bf Proof.}
We can first evaluate  the self-energy \eqref{topt} as 
\beq
\Sigma(b_1,b_2, b_3) =  \sum_{\cG }  K_{\cG}\, A_{\cG}(b_1,b_2, b_3),
\label{sigma}
\eeq
where $\cG$ runs over two-point 1PI graphs 
with amplitude $A_{\cG}(b_1,b_2, b_3)$
and corresponding combinatorial weight $K_{\cG}$.  
The latter is known to be the number of Wick contractions given rise to 
$\cG$. 
At one-loop, only the tadpole graphs $T_1$, $T_2$, $T_3$ 
(such that $\sum_{J} g_{\tJ}=0$) 
$T'_1$, $T'_2$ and $T'_3$ (such that $\sum_{J} g_{\tJ}=1$) 
(Fig.\ref{fig:tadpoles} provides $T_1$ and $T_1'$, the remaining configurations can be easily recovered by color permutations   
corresponding to the other configurations of the vertex $\varphi^4$) contribute to (\ref{sigma}) with the combinatorial factors
\beq
K_{T_\epsilon}=2 =K_{T'_\epsilon} ,\qquad \epsilon= 1,2,3.
\eeq  
One gets the amplitude for each tadpole $T_{\epsilon}$
\beq
A_{T_\epsilon}(b_\epsilon)=-\frac{\lambda_\epsilon}{2}\mathcal{S}_\epsilon(b_\epsilon),
\eeq
where $\mathcal{S}_\epsilon(b_\epsilon)$ is given by \eqref{sepbe}.
Meanwhile, for $T'_{\epsilon}$, we have the amplitude
\beq
A_{T'_\epsilon}(b_{\check \epsilon}, b_{\check{\check\epsilon}} )=-\frac{\lambda_1}{2}\mathcal{S}'_\epsilon( b_{\check \epsilon}, b_{\check{\check\epsilon}} )   
\eeq
with 
$\mathcal{S}'_\epsilon( b_{\check \epsilon}, b_{\check{\check\epsilon}} ) $ given as \eqref{seppbe}.
At first order, we obtain the self-energy as the sum of these contributions as
\bea
\Sigma(b_1,b_2,b_3) =  - \sum_{\epsilon} \lambda_{\epsilon} 
\mathcal{S}_\epsilon(b_\epsilon)  -\lambda_1\mathcal{S}'_1( b_{2}, b_{3} )   
-\lambda_2\mathcal{S}'_2( b_{1}, b_{3} )   
-\lambda_3\mathcal{S}'_3( b_{1}, b_{2} ) 
+O(\lambda^2) .
\label{sigmafin}
\eea
It should be emphasized that not all contributions of $\Sigma(b_1,b_2,b_3)$ have to be taken into account for 
the wave function renormalizations \eqref{wfr}.   Only those
leading to a log-divergent behavior after differentiation have to be considered. 
A quick inspection shows that $\mathcal{S}'_\epsilon(b,b')$ 
are log-divergent hence should be only considered for the  
mass renormalization.
The wave function renormalizations \eqref{wfr} can be finally expressed as
\beq
Z_{\epsilon}= 1-  \lambda_\epsilon\,  S_{\epsilon} 
+O(\lambda^2).
\label{sigz}
\eeq
where $S_{\epsilon}$ is given by \eqref{sep}. 
\qed

We are in position to compute the dynamics of the
renormalized wave coupling constants $a^\ren_\epsilon$ 
and of the renormalized mass $m^\ren$. In the same previous notations,
the following statement holds:
\begin{theorem}
\label{theo:wrfandmass}  
At first order, 
the renormalized wave function couplings and mass satisfy, respectively,
\bea
&&
a^\ren_\epsilon =  a_{\epsilon}  \left[ 1 - \frac13\big(2 \lambda_\epsilon
 S_{\epsilon}-\lambda_{\check\epsilon} S_{\check\epsilon} \, -\lambda_{\check{\check\epsilon}} 
 S_{ \check{\check\epsilon} }\big) \right]   
+O(\lambda^2),\; \quad \epsilon = 1,2,3, \;
\check \epsilon \neq \epsilon, \;
\check \epsilon \neq \check{\check \epsilon} 
\neq  \epsilon ,
\label{aren}\\
&&
m^\ren  =  m + \sum_{\epsilon=1}^3 \lambda_{\epsilon}\mathcal S_\epsilon(0)   
+O(\lambda^2,\lambda\ln \Lambda).
\label{massren}
\eea
 for momentum cut-off $\Lambda$.
\end{theorem}
\noindent{\bf Proof.} Concerning the first statement \eqref{aren}, by simple index permutations, all cases can be easily deduced from $\epsilon = 1$ 
hence we will focus only on this situation.  Using Lemma \ref{self} 
and remaining at first order in the constant couplings $\lambda_\epsilon$, we have:
\beq
a_{1}^{\ren} 
=
   a_{1}  \left(\frac{ 1- 2\lambda_1\,  S_1  
 +O(\lambda^2)
}{1- (\lambda_2 S_2\, + \lambda_3 S_3)
+O(\lambda^2)}
\right)^{\frac13} 
 =  a_{1}  \left[1  - \frac13\Big(2 \lambda_1 S_1 
- \lambda_2 S_2\, -  \lambda_3 S_3\Big)
\right]  
+O(\lambda^2).
\eeq
Focusing on the renormalized mass equation \eqref{mreninit}, Lemma \ref{self} allows us to write
\bea
m^\ren &=&  \frac{m +\sum_{\epsilon} \lambda_{\epsilon} 
\mathcal{S}_{\epsilon} (0)  +\lambda_1\mathcal{S}'_1( 0,0 )   
+\lambda_2\mathcal{S}'_2( 0,0 )   
+\lambda_3\mathcal{S}'_3( 0,0 ) + 
O(\lambda^2)}
{1- \frac13\left( \lambda_1 S_1
+ \lambda_2 S_2+\lambda_3\,  S_3\right)+
O(\lambda^2)} \crcr
& =&
 m + \sum_{\epsilon} \lambda_{\epsilon} 
\mathcal{S}_{\epsilon} (0)  +\lambda_1\mathcal{S}'_1( 0,0 )   
+\lambda_2\mathcal{S}'_2( 0,0 )   
+\lambda_3\mathcal{S}'_3( 0,0 ) \crcr
&&
 + \frac13\left( \lambda_1 S_1
+ \lambda_2 S_2+\lambda_3\,  S_3\right)   
+  O(\lambda^2).
 \eea
Neglecting the subleading divergences compared to the linear 
divergence of $\mathcal{S}_{\epsilon} (0)$, one is led to \eqref{massren} 
after having introduced a cut-off $\Lambda$ in the sums 
$\mathcal{S}'_\epsilon(0,0)$ and $ S_\epsilon$. 
\qed

\noindent{\bf Discussion.}
The $\gamma$-functions of
the model can be defined by restricting the space of the couplings
parameters to a smaller subspace. In fact, the reduction
 will be performed on the space of six couplings $\lambda_{1,2,3}$ and $a_{1,2,3}.$

Many cases may occur by collapsing
couplings. The first case is straightforward

\begin{enumerate}

\item[(i)]  We set $a_\epsilon =a$  in \eqref{aren}. This enforces $\lambda_1 = \lambda_2 = \lambda_3$ and induces
a unique equation:
\bea
a^\ren =  a .
\label{arena}
\eea
The latter means that the wave function coupling becomes stationary
in the UV. In other 
words, the $\gamma$-function for this case is trivial: 
\beq
\gamma = 0.
\eeq
Meanwhile, from \eqref{massren}, one infers 
\beq
\beta_m = -3.
\eeq

\item[(ii)] Otherwise, setting $\lambda_2=\lambda_3=0$ but 
$\lambda_1 \neq 0$  yields necessarily that $a_1 \neq a_2$. 
We need to re-evaluate the self-energy from \eqref{sigmafin}
and we obtain  the only non trivial wave function renormalization
\beq
Z^{(')}_1 = 1 - \lambda_1 S'_1  +O(\lambda_1^2) .
\eeq
where $S'_1$ is given by \eqref{sprim}. 
The equations of the couplings find some modified form:
\beq
a_1^{(')\,\ren} = a_1,\qquad  a_2^{(')\,\ren} = a_2 \frac{1}{Z_1}
 = a_2 (1+  \lambda_1 S'_1) + O(\lambda_1^2)\,.
\label{arena1}
\eeq
The $\gamma$-functions at one-loop in this restricted space
  are given by 
\beq
\gamma^{(')}_1 = 0 , \qquad \gamma^{(')}_2 = -1.
\eeq
Assuming that the sign of $\lambda_1$ is positive, 
$a_1$ has a stable value in the UV, whereas $a_2$
flows towards a  vanishing value.  Under the same 
assumptions, we get for the mass, the renormalized
mass equation and $\beta_m$ function
\beq
m^{(')\,\ren} = m + \lambda_1 \tilde{\mathcal S}_1(0) + O(\lambda_1^2,\lambda_1 \ln \Lambda) ,\qquad
\beta^{(')}_m = -1,
\eeq
where $\tilde{\mathcal S}_1(0)$ can be inferred from 
${\mathcal S}_1(0)$  after identifying $a_2=a_3$.

\item[(iii)] Setting $\lambda_2=\lambda_3\neq 0$,
$\lambda_1 = 0$ and $a_1 \neq a_2$ and computing 
the self-energy, one has 
the unique non trivial wave function renormalization 
\beq
Z_2^{('')} = 1 - \lambda_2 S'_2  + O( \lambda_2^2),
\eeq
where $S'_2$ is provided by \eqref{sprim}. 
In this restricted space,  the coupling equations  are given by 
\beq
a_2^{('')\, \ren} = a_2 , 
\qquad 
a_1^{('')\,\ren} = a_1 \frac{1}{Z_2^{('')}} =a_1 (1+\lambda_2 S'_2)  + O( \lambda_2^2) , 
\eeq
yielding the $\gamma$-functions at one-loop  
\beq
\gamma_2^{('')} = 0 , \qquad \gamma_1^{('')} = -1.
\eeq
The same  assumptions give for the mass
\beq
\beta^{('')}_m =-1.
\eeq
Thus cases (ii) and (iii) share  similar properties.

\end{enumerate}

 \begin{theorem}
\label{theo}
The renormalizable coupling constants  associated with $\lambda_\epsilon,$ for $\epsilon=1,2,3,$ satisfy
\beq
\lambda_{\epsilon}^{\ren} = 
\lambda_\epsilon 
+ \lambda_\epsilon\left[-\lambda_\epsilon S_\epsilon 
 +\frac{2}{3}\Big( \lambda_1 S_1+\lambda_2 S_2+\lambda_3 S_3 \Big)   \right]
+O(\lambda^3),
\label{lamere}
\eeq
where $O(\lambda^3)$  is a O-function of any cubic product of 
all couplings $\lambda_{1,2,3}$. 
\end{theorem} 
\noindent {\bf Proof.} The one-loop 1PI four-point functions 
can be written as
\beq
\Gamma_{4,\epsilon}(b_1,b_2, b_3,b_1',b_2', b_3')= \sum_{\cG} K_{\cG}\, \mathcal{A}_{\cG}(b_1,b_2, b_3,b_1',b_2', b_3')
\label{gamm4}
\eeq
where $\cG$  is a four-point 1PI graph with the topology as required by the renormalization, with
amplitude $\mathcal{A}_{\cG }(b_1,b_2, b_3,b_1',b_2', b_3')$ 
and combinatorial weight $K_{\cG}$. 

\begin{figure}
 \centering
    \begin{minipage}[t]{.7\textwidth}
     \centering
 \includegraphics[angle=0, width=6cm, height=2cm]{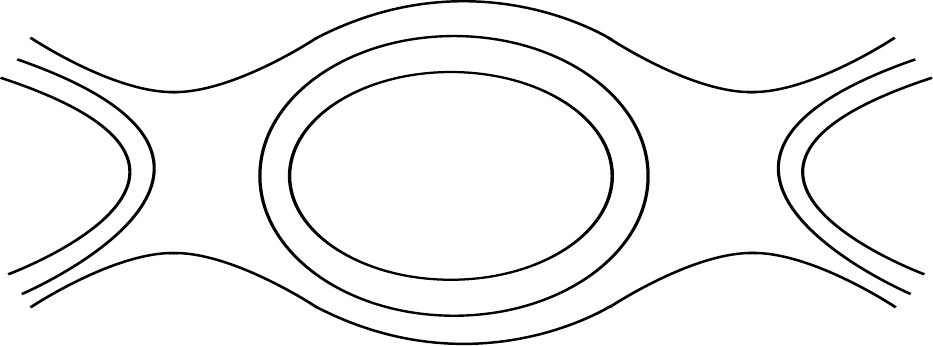}
 \caption{ {\small The unique melonic 1-loop four-point function 
with melonic connected boundary for the first type of vertex. }}
\label{fig:fourpt}
 \end{minipage}
 \end{figure}

Note that, interestingly, the ``mixed'' graphs obtained by gluing 
vertices of different kind  do not contribute
to the effective coupling constants. Indeed, although these
graphs could form melons, they all possess two lines and one
face giving a convergent power-counting $M^{-1}$. Another way to 
agree with this fact is to observe that the boundary graph 
(even though melonic) of any of them is disconnected with $C_{\bG}=2$.
As a consequence, in the following developments, cross terms 
involving product of coupling constants $\lambda_\epsilon\lambda_{\epsilon'},$ $\epsilon\neq \epsilon',$ are inexistent. Hence, a unique graph $F_{\epsilon}$ contribute to $\Gamma_{4,\epsilon}$  (Fig.\ref{fig:fourpt} displays such a 
contribution for $\epsilon=1$, the remaining can be deduced 
by permutations) and its combinatorial factor is always
\beq
K_{F_\epsilon}=2 \cdot 2 \cdot 2  \,.
\eeq
Given an $\epsilon$, the amplitude of each graph is such that 
\bea
{\mathcal A}_{F_\epsilon}(b_\epsilon,b_\epsilon')& =&\frac{\lambda^2_\epsilon}{2^2 2!}\,\underline{\mathcal S}_\epsilon(b_\epsilon,b_\epsilon'),
\eea
where $\underline{\mathcal S}_\epsilon(b_\epsilon,b_\epsilon')$
is given by \eqref{undes}.

The amputated 1PI four-point functions $\Gamma_{4,\epsilon=1,2,3 }$ defined by \eqref{fopt} and \eqref{gamm4} can be evaluated at one-loop as 
\beq
\Gamma_{4,\epsilon}(b_1,b_2, b_3,b_1',b_2', b_3')= - \lambda_\epsilon +\lambda_\epsilon^2\,\underline{\mathcal S}_\epsilon(b_\epsilon,b_\epsilon') 
+ O(\lambda^3),
\eeq 
where $O(\lambda^3)$ is a O-function of any cubic
product of coupling constants.
Therefore, at low external momenta, they reduce to
\beq
\Gamma_{4,\epsilon}(0,0,0,0,0,0) = - \lambda_\epsilon +
\lambda_\epsilon^2 \,S_\epsilon 
+ O(\lambda^3).
\label{gamresult}
\eeq
Having all required ingredients, we are in position to evaluate the
ratios \eqref{gamma4}. Using Lemma \ref{self} and \eqref{gamresult},
 by direct algebra, one comes to
\begin{eqnarray}
\label{lamrenep}
\lambda_{\epsilon}^\ren&=&-\frac{ - \lambda_\epsilon +\lambda_\epsilon^2 S_\epsilon+O(\lambda^3)}{
\left\{
\prod_{\epsilon}
\left(1- \lambda_{\epsilon}\, S_\epsilon + O(\lambda^2) \right)\right\}^{\frac23}}\crcr
&=& 
\lambda_\epsilon 
+ \lambda_\epsilon\Bigg(-\lambda_\epsilon S_\epsilon 
 +\frac{2}{3}\Big( \lambda_1 S_1+\lambda_2 S_2+\lambda_3 S_3 \Big)   \Bigg)  
+O(\lambda^3)
\end{eqnarray}
which achieves the proof.
\qed
\medskip 

\noindent{\bf Discussion.} We can once again discuss the merging of coupling constants $\lambda_{1,2,3}$ 
into specific varieties in order to deduce the $\beta$-functions.

\begin{enumerate}

\item[(i)]  We merge all couplings such that $a_{\epsilon}=a$ and $\lambda_\epsilon=\lambda$, such that $S_{\epsilon}$ \eqref{sep} becomes 
\beq
S = \sum_{p_{1},p_{2}\in \Z} 1/[a(|p_{1}|+|p_2|)+m)]^2,
\label{sasum}
\eeq
then we have from \eqref{lamere}
\beq
\lambda^\ren =
\lambda + \lambda^2 S +O(\lambda^3). 
\label{larenla}
\eeq
such that the $\beta$-function of the model 
with single wave function renormalization and single
coupling constant is given by
\beq
\beta = -1.
\eeq
This model is therefore  asymptotically free.

\item[(ii)] Assuming that $a_1 \neq a_{2}=a_3$, 
and $\lambda_1 \neq 0 = \lambda_{2,3}$, there
is a unique equation for the coupling constant $\lambda_1$
expressed as
\beq
\lambda^{(')\,\ren}_1 = - \frac{\Gamma^{(')}_{4,1}}{(Z_1^{(')})^2}
 = - \frac{(-\lambda_1  + \lambda_1^2 S_1'  + O(\lambda^3_1)))}{(1- \lambda_1 S_1' + O(\lambda^2_1))^2} 
 = 
\lambda_1 
 +\lambda_1^2 S_1' +O(\lambda_1^3) .
\label{larenlab}
\eeq
From this, we infer
\bea
\beta^{(')}_1 = -1.
\eea
Consequently,  the model is also asymptotically free.  

\item[(iii)] Last, we set $\lambda_2=\lambda_3\neq 0$, 
$\lambda_1 = 0$ and $a_1 \neq a_2$. 
The coupling constant equation can be written 
\beq
\lambda_2^{('')\, \ren} =  - \frac{\Gamma^{('')}_{4,2}}{(Z_2^{('')})^2}
\eeq
and yields in the same way as previously done in \eqref{larenlab}
the one-loop $\beta$-function 
\beq
\beta^{('')}_{2} = -1  
\eeq
showing that this theory is asymptotically free. 

\end{enumerate}

\subsection{RG flows}
\label{subsect:flow}

We restrict the RG flow equations to the particular situations
discussed so far. It turns out that for these cases, explicit 
solutions are affordable at this truncation. Cases (i) and (ii) will be only discussed
for simplicity.

\begin{figure}
 \centering
     \begin{minipage}[t]{.8\textwidth}
      \centering
\includegraphics[angle=0, width=6.5cm, height=4cm]{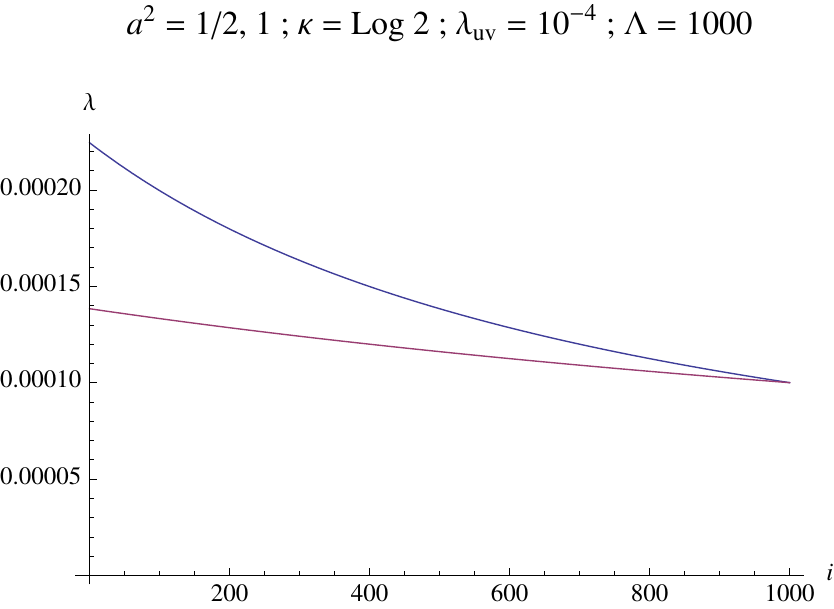}
\caption{ {\small RG flow of the coupling constant $\lambda$
running to a trivial fixed point for two values $a=1/2$ (top) and $a=1$ (bottom) and for 
a UV cut-off $\Lambda=1000$ and other parameters fixed such that $\lambda_{uv}=10^{-4}$ and $\kappa=\log M=\log 2$.}}
\label{fig:rg1}
\end{minipage}
\end{figure}

\noindent{\bf Case (i)}: $\lambda_\epsilon= \lambda$ and $a_\epsilon=a$. This case yields 
the coupling equations \eqref{arena} and \eqref{larenla} of the form:
\bea
a_\epsilon^\ren = a, \qquad 
\lambda^\ren = \lambda + \lambda^2 S + O(\lambda^3).
\eea
Hence, the wave function renormalization is stable 
whereas the coupling constant satisfies the discrete RG equation
(truncated at first order)
\beq
\lambda_{i-1} = \lambda_i + \lambda_i^2\, S_i,
\eeq
where $S_i$ is given from $S$ \eqref{sasum} by  restricting the 
sum to high momenta $|p_1|+|p_2| \in [M^{i-1},M^{i}]$ such that (see Appendix \ref{app:formal} for more details)
\beq
S_{i} 
 = \frac{4}{a^2_{i}} \,\kappa_M + O(M^{-i}),
\label{sum2}
\eeq
with $\kappa_M$ a log-divergent term in $M$
which does not depend on the scale $i$. We get the 
solution, using $a_i =a$,
\beq
\frac{d\lambda_i}{\lambda_i^2} = -\frac{4}{a^2} \,\kappa_M \,di, 
\qquad 
\lambda(i)  = \frac{a^2 \lambda_{uv}}{a^2 - 4 \kappa_M (\Lambda - i) \lambda_{uv}},
\eeq
where we have introduced the coupling value $\lambda_{uv}$ at the
cut-off scale $\Lambda$. The RG flow of $\lambda$ is pictured
in Fig.\ref{fig:rg1}.

\medskip

\noindent{\bf Case (ii)}: $\lambda_2 = \lambda_3 =0 \neq \lambda_1$ and $a_2=a_3\neq a_1$.
The RG flow equations can be obtained from \eqref{arena1} and   \eqref{larenlab} as 
\bea
&&
a_{1}^\ren= a_1  ,\qquad
a_{2}^\ren=   a_{2}  (1+\lambda_1 S'_1) + O(\lambda_1^2), 
\crcr
&&
\lambda_{1}^\ren=  \lambda_{1} +  \lambda_{1}^2\, S_1'  + O(\lambda_1^3), 
\eea
with $S_1'$ given by \eqref{sprim}.
The same RG flow equations (truncated at one loop corrections) translate in discrete version at scale $i$  as 
\beq
a_{1;\;i-1}=   a_{1;\;i}   ,\;\;
a_{2;\;i-1}=   a_{2;\;i}   \big[1  +\lambda_{1;\; i}\, S_{1,\;i}' \big], \;\;
\lambda_{1;\;i-1}=  \lambda_{1;\;i}+ \lambda_{1;\;i} ^2\,S'_{1;\;i} .
\eeq
Summing $S_1'$ only on high momenta $|p_1|+|p_2| \in [M^{i-1},M^{i}]$, 
we have
$
S'_{1;\;i} = 4/a^2_{2;\;i}\kappa_M + O(M^{-i})$.
Focusing of the second and third expressions, these discrete equations  can be put, at first order, in an equivalent differential form
\bea
\frac{d a_{2;\;i}}{di} &=& -4\,\frac{1}{a_{2;\;i}}\,  \lambda_{1;\; i} \,\kappa_M 
\crcr
\frac{d\lambda_{1;\;i}}{di} &=&-4\,\frac{1}{a^2_{2;\;i}}\,  \lambda_{1;\;i} ^2\,
\,\kappa_M 
\label{diffeq}
\eea
which induces that
\bea
\lambda_{1;\;i}\, /a_{2;\;i} =\tilde K,
\label{l1ak}
\eea
for some constant $\tilde K$. We substitute $\tilde K $ in \eqref{diffeq} and get 
\bea
d\lambda_{1;\;i} = - 4\kappa_M \tilde K^2 \, di 
\quad   \Leftrightarrow \quad  
\lambda_1(i)  =  \lambda_{uv} + 4\kappa_M \tilde K^2 (\Lambda - i).
\label{l1i}
\eea
Thus, one can readily obtain $a_2(i)$ by combining \eqref{l1i} and
\eqref{l1ak}. 
We see that the flow of this model actually differs from the
previous one. Both couplings $\lambda_1$ and $a_2$ go linearly to their 
UV fixed values.

\section{Conclusion}

We have shown that a rank 3 tensor model on $U(1)^3$,
analogue of the previous model identified in \cite{arXiv:1111.4997}, 
is perturbatively renormalizable at all orders of perturbation theory.
The  proof of this statement relies on both combinatorics and the
colored model properties serving as underpinning of the
present class of tensor models. 
The interplay of three ingredients, 
namely a slice decomposition, a power-counting theorem
and a generalized locality principle has guided us towards a 
proof of renormalizability of this model. 
We have also introduced different wave function couplings for each 
strand hence giving them a different dynamics 
as well as different interaction coupling constants. 
This definitely enlarges the class of renormalizable models
of the kind. As it should be also emphasized, the renormalization
has been performed in the momentum space and not
in the direct space. This particular 
basis has allowed us to identify an analogue theorem for the momentum 
routine useful for  renormalization in ordinary quantum field theory \cite{Rivasseau:1991ub}. The second part of this contribution 
was devoted to the computations of the $\gamma$-, $\beta$-functions
and the corresponding RG flow of couplings constants in 
the theory.  
There exist two classes of 
underlying models  turn out to be asymptotically free in the UV limit: 
(a) The model obtained by merging all coupling constants, such that $\lambda_{1,2,3} =\lambda$ and $a_{1,2,3}=a$; 
 (b) the class of models defined by $\lambda_{2,3} =0 \neq \lambda_1$
and $a_{1}\neq a_2 = a_3$ and the symmetric obtained by color
permutation. 

The  model introduced here and its analogue developed
in \cite{arXiv:1111.4997} claimed to be simplified but 
possible models for quantum gravity using the ``tensor device'' \cite{Rivasseau:2011hm}. What we have shown here, by exhibiting such a
renormalizable and asymptotically free (in the UV) model, 
is that the tensor approach for quantum gravity is a 
promising line of research for those who believe
that gravity should be described by a renormalizable 
quantum field theory.

Future prospects are can be performed in different directions. 
Nonlocal tensor models over copies of $\mathbb{R}$ can be treated
provided one introduces an IR regulator. For instance, the Mehler
propagator $1/(p^2 + x^2)$ (on each strand) could be an interesting proposal. The resulting models should be called tensor extension of 
the Grosse-Wulkenhaar model \cite{Grosse:2004yu} 
in noncommutative field theory.
Another important question is whether or not such an analysis could be
handled for GFT models on copies of $SU(2)$. A priori, since the tensor $1/N$ expansion is also valid for GFT models, the answer is yes. Nevertheless, the level of difficulty is much higher is that situation.
The renormalization will definitely involve both $SU(2)$ recoupling theory and Taylor expansions of graphs around their local parts. These later entail generalized saddle point analysis which are far to be easy for general graphs at any order of perturbation theory 
\cite{Geloun:2011cy}. 
Besides, going deeper in the analysis of the class of models presented
here, an important point would be to investigate if the rank four
model shares this important property of asymptotic freedom.
Finally, interesting though more abstract questions can be also 
addressed at this stage. For instance, the determination of topological polynomials associated with the graphs of these models using an extended (i.e. tensor) parametric representation of graph 
amplitudes in the spirit of \cite{Krajewski:2010pt} could be
a fruitful line of investigation.

\section*{ Appendix}
\label{app}

\appendix

\renewcommand{\theequation}{\Alph{section}.\arabic{equation}}
\setcounter{equation}{0}

\section{Proof of Lemma \ref{localjacket} 
and divergent graph classification}
\label{app:lemclass}

\subsection{Proof of Lemma \ref{localjacket} }
\label{app:prooflem}

\begin{figure}
 \centering
    \begin{minipage}[t]{.7\textwidth}
      \centering
 \includegraphics[angle=0, width=11cm, height=4cm]{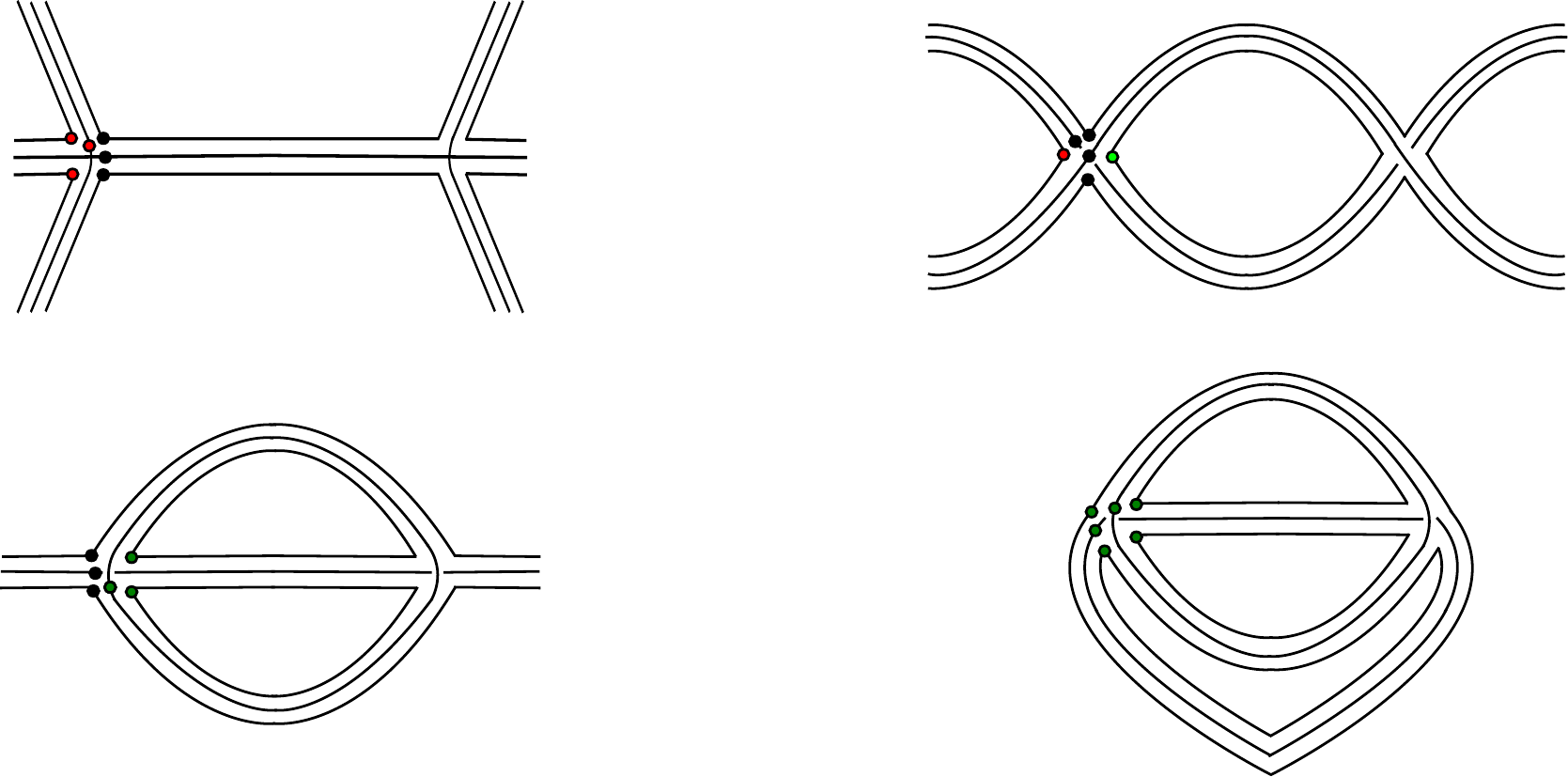}
 \caption{ {\small Different types of pairs highlighted for each $0k$-dipoles: External pairs (in red), internal pairs (in green)
and mixed pairs (in black).   }}
\label{fig:pairs}
\end{minipage}
 \end{figure}

This appendix provides the proof of Lemma \ref{localjacket}
using a sequence of $0k$-dipole contractions. 
First of all, we need to define a central notion which is the following:

\begin{definition}[Pairs] Let $\cexG$ be a colored graph 
and consider a $0k$-dipole inside $\cexG$. 
A ``pair'' is a couple of colors $(ab)$, $a,b=0,1,2,3$. 
A pair is called ``outer'' if the two colors are external to the dipole. 
A pair which has one color inside the dipole and one out is called a ``mixed'' pair. 
A pair with two colors inside the dipole is called an ``inner'' pair.
A pair $(ab)$ is said to belong to a jacket if the pair is one of the four adjacent (consecutive) pairs in the jacket cycle $(0xyz)$. 
\end{definition}
The total number of pairs is always 6 and the number of mixed pairs is at least 3. Fig.\ref{fig:pairs} indicates the different kinds of pairs.

Contracting a dipole leads to different cases and we urge to 
further classify the different pairs in a more concise manner. 
An outer pair is said of type A, or disconnected by the dipole contraction
if the half-strands at each corner on the left and on the right of the dipole belong to two different
connected components of the graph after the contraction. Otherwise, we call it a ``special'' pair. A special 
pair can be single-faced if the two corners belong to the same face of the graph,
or double-faced if the two corners  belong to two different faces of the graph. Any type A outer pair 
should be single-faced at the beginning. 
We therefore classify outer pairs in the following three types:
\begin{itemize}
\item Type A outer pairs are single-faced,
\item Type B outer pairs are single-faced,
\item Type C outer pairs are double-faced.
\end{itemize}

Transverse pairs are mixed pairs which do not change their number of faces under contraction. Inner pairs have one face less
after contraction. Type A and B outer pairs have one face more after contraction whereas type C external pairs
have one face less after the contraction.
In summary, for any jacket $\tJ$, we can relate the number
of faces before and after contraction by the formula
\beq
F_{\tJ' } - F_{\tJ}  =  \vert A_{\tJ} \vert + \vert B_{\tJ} \vert - \vert C_{\tJ} \vert - \vert I_{\tJ} \vert \;,
\eeq
where $ \vert X_{\tJ} \vert $ is the number of pairs of type X=A,B,C, in the jacket, and $|I_{\tJ}|$ is the number of inner faces.

Lemma \ref{localjacket} states that,  for a single jacket $\tJ$, we have 
\beq
 g_{\tJ} - g_{\tJ'}  \geq  0 .
\label{gtgj}
\eeq
We emphasize that the proof of this statement which will be given 
here slightly differs from the proof of a related statement as found in \cite{arXiv:1111.4997}. Indeed, in the latter reference, the sum over jackets is performed at the level of \eqref{gtgj} and leads directly 
to a bound on $ \sum_{j}(g_{\tJ} - g_{\tJ'}) $. However, as was 
mentioned \cite{arXiv:1111.4997} and was computed
 in a earlier version of the same paper, the relation \eqref{gtgj}
is a stronger statement than the result on the sum of jackets
and therefore worth to be communicated.

\noindent{\bf Proof of Lemma \ref{localjacket}.}
 We first translate the difference between genera 
of jackets before and after contraction in terms 
of the basic combinatorial elements:
\bea
2 - 2g_{\tJ} &=&  V - L+ F_{\tJ}\;, \crcr
2 c'- 2g_{\tJ' } &=&V' - L'+ F_{\tJ' } = (V-2) - (L-4) + F_{\tJ' }  \;,\crcr
(g_{\tJ} - g_{\tJ'})  & =& \frac{1}{2}[ (F_{\tJ' } - F_{\tJ} )  +2  - 2(c'-1)]\;.
\eea
The sign of $g_{\tJ} - g_{\tJ'}$ reduces to 
that of $ (F_{\tJ' } - F_{\tJ} )  +2  - 2(c'-1)$. The latter quantity proves to be always positive jacket by jacket. To prove this claim,
we perform the following systematic analysis on the types of 
$0k$-dipole contractions.

\medskip
\noindent{ $\bullet$ \bf 1rst Case:  00-dipole contraction.} This case is defined by a contraction of a unique internal line with color $0$. 
In any initial colored graph, there are three mixed pairs, three outer pairs and no inner pair. Meanwhile, each jacket contains two mixed and two outer pairs.

\begin{figure}
 \centering
    \begin{minipage}[t]{.7\textwidth}
     \centering
 \includegraphics[angle=0, width=10cm, height=2cm]{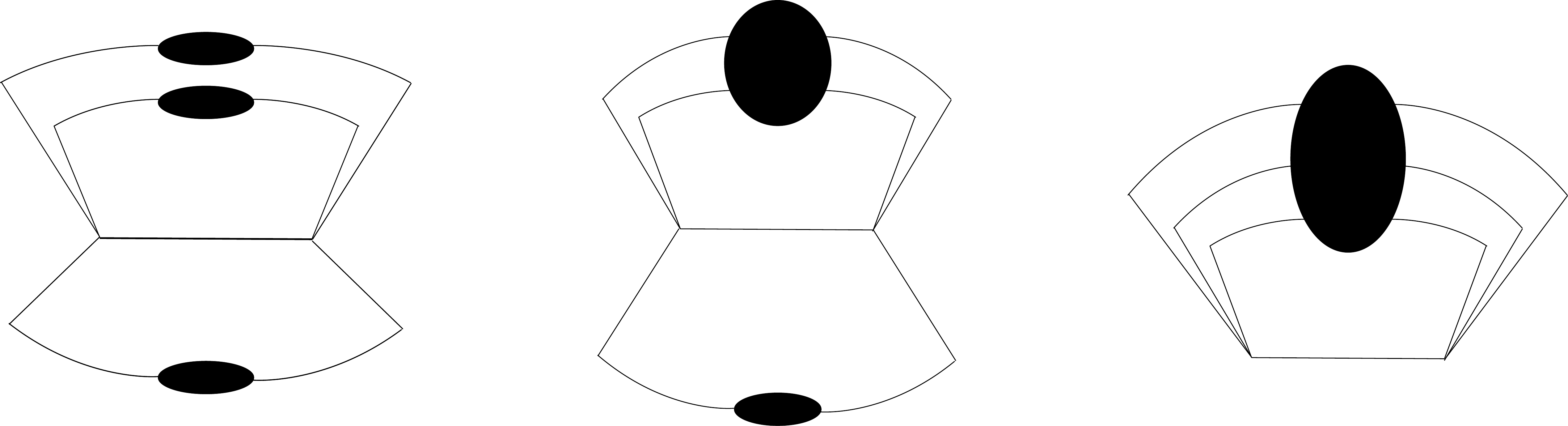}
 \vspace{0.1cm}
 \caption{ {\small $00$-dipoles configurations.}}
\label{fig:dipcont}
 \end{minipage}
 \put(-275,-12){A}
 \put(-170,-12){B}
 \put(-67,-12){C}
 \end{figure}

\noindent
{\it - 1rst subcase $c'=3$ (Fig.\ref{fig:dipcont}A)}.  
This situation happens if the initial graph has a connected two-point subgraph on each line $1,2,3$.
Consequently, all three outer pairs must be of type $A$.  
Hence, the $00$-dipole contraction yields for all jackets:  
\beq
(F_{\tJ' } - F_{\tJ} )  +2  - 2(c'-1) =0.
\eeq
so that \eqref{interm1} is true.

\noindent
{\it - 2nd subcase  $c'=2$ (Fig.\ref{fig:dipcont}B)}. This case happens if we have one connected two-point 
functions plus one connected four-point function
on four half-lines hooked to the dipole. In that case, we have
two corner pairs of type A and one special pair of type B or  C. 
It can be seen that
\beq  \label{special1}
(F_{\tJ' } - F_{\tJ} ) + 2  - 2(c'-1)  = 2 - 2 \vert C_{\tJ}  \vert .
\eeq
Hence in all cases, \eqref{interm1} is true.

\noindent
{\it - 3rd subcase $c'=1$ (Fig.\ref{fig:dipcont}C)}. Contracting the dipole 
gives $c'-c=0$. This can happen when 
a connected six-point function is hooked to 
the dipole. There no type A corners and three special 
pairs. Each jackets contains at most two special pairs.
Contracting the dipole yields
\beq
(F_{\tJ' } - F_{\tJ} ) + 2  - 2(c'-1)  = 4 - 2 \vert C_{\tJ}  \vert ,
\eeq
such that 
 \eqref{interm1} 
is  true.

\medskip
\noindent{ $\bullet$ \bf 2nd Case:  01-dipole contraction.}
There is one inner, four mixed and one outer pairs.
A distinction should be made between 
inner-adjacent jackets, for which the 
inner pair of the dipole belongs to the jacket 
(see Fig.\ref{fig:inner}A, for instance,
the pair $(03)$ belongs both to the jacket $(0123)$ and the dipole) 
from non-inner-adjacent jackets for which it does not (see Fig.\ref{fig:inner}B, for instance, the pair $(02)$ is not in the jacket $(0123)$ but defines
the dipole). We count two subcases:

\begin{figure}
 \centering
    \begin{minipage}[t]{.7\textwidth}
      \centering
\includegraphics[angle=0, width=8cm, height=2cm]{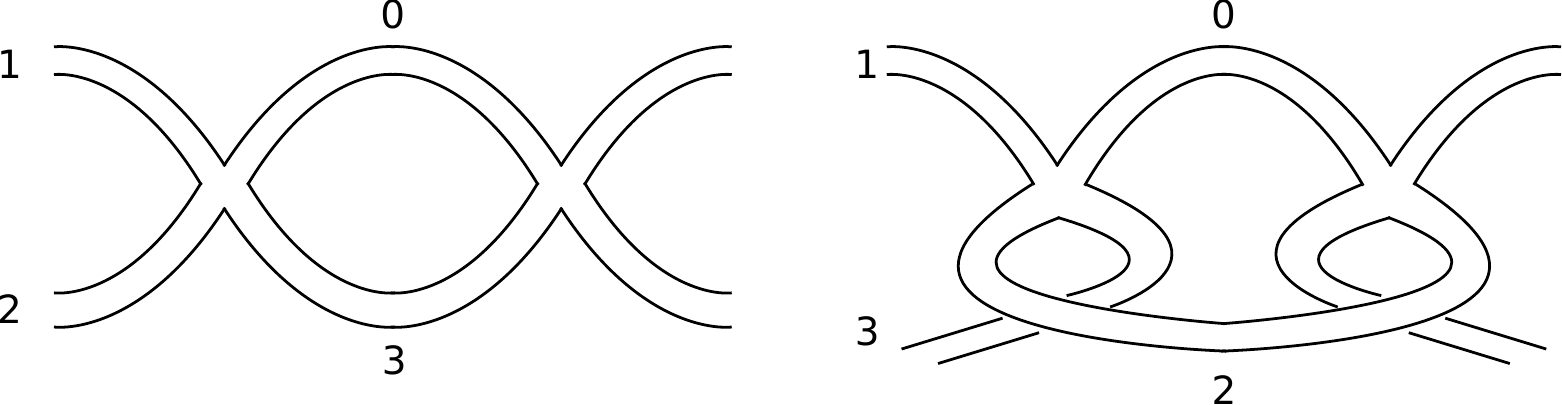}
\vspace{0.1cm} 
\caption{ {\small A jacket $(0123)$ which contains the pair $(03)$ 
defining the dipole (A) and another jacket $(0123)$ from another dipole 
which does not contain the same pair. }}
\label{fig:inner}
 \end{minipage}
 \put(-225,-10){A}
 \put(-102,-10){B}
 \end{figure}

\noindent{\it - 1rst subcase $c'=2$}. In that case, the unique outer pair is of type A.
An inner-adjacent jacket has two mixed pairs, one inner pair and one pair of type A. 
A non-inner-adjacent jacket has four mixed pairs, no inner pair and 
looses the type A pair. One finds, in all cases,
\beq
(F_{\tJ' } - F_{\tJ} ) + 2  - 2(c'-1)  = 4 -4 =0,
\eeq
and \eqref{interm1} 
is verified.

\noindent{\it - 2nd subcase  $c'=1$}.
This case reduces to the presence of one  special pair.
An inner-adjacent jacket has two mixed pairs, one inner pair and one special. A non-inner-adjacent jacket has four mixed pairs, no inner pair and
no special pair. In all cases,
\beq
(F_{\tJ' } - F_{\tJ} ) + 3  - 2(c'-1)  = 2 - 2 \vert C_{\tJ}  \vert 
\eeq
and \eqref{interm1} is true. 

\medskip
\noindent{ $\bullet$ \bf  3rd Case:  02-dipole contraction.}
There are three inner pairs, three mixed pairs and no outer pair. 
Only the case $c'=1$ occurs and one gets
\beq
(F_{\tJ' } - F_{\tJ} ) + 2  - 2(c'-1)  =0,
\eeq
hence, \eqref{interm1} 
is satisfied.

\medskip
\noindent{ $\bullet$ \bf  4th Case: 03-dipole contraction.}  
This case is straightforward and corresponds to 
the contraction of a vacuum connected component
(with two vertices and four lines).
There are six inner pairs, each jacket containing four of
these pairs and we have $ (c'-1)  = -1$. Thus
\beq
(F_{\tJ' } - F_{\tJ} ) + 2  - 2(c'-1)  = 0 \;.
\eeq
Hence \eqref{interm1}  
is again valid.

\qed

\subsection{Classification of divergent graphs}
\label{app:conv}

We extract from the divergence degree data all 
graphs which could lead to $\omega_d(\cG)\geq 0$
and aim at proving that Table 1 of Subsection \ref{subsect:digraph}
exhausts all possibilities. 

Let us define the integer $P(\cG)= (C_{\bG}-1)  
+V_2+ \frac12  \left[ N_{\ext} - 4\right]$.
We reformulate Lemma \ref{genbound} in the following terms:
\bea 
\label{betterbound1}
\omega_d(\cG) = -\Big[ 
\sum_{J} g_{\tJ} - g_{\bG} \Big]-P(\cG) &\leq&   - 2g_{\bG}   -P(\cG)  ,  \\
 g_{\bG} =0\;\; {\rm and} \;\; \sum_{J} g_{\tJ} >0 \quad & \Rightarrow& \quad \omega_d(\cG)
\leq -1  -P(\cG) \,   . 
\label{betterbound3}
\eea
We now seek the list of graphs with  $\omega_d(\cG)  \geq 0$ which are
those which should be renormalized.

\noindent{\bf Case} $N_{\ext} > 4$: 
Considering $N_{\ext}\geq 6$ leads to $P(\cG) \geq 1$.
Thus, $\omega(\cG) \leq -1$ and the graph amplitude
is simply converging. 

\noindent{\bf Case}  $N_{\ext} = 4$:
Under this condition, one has $P(\cG)= (C_{\bG}-1) +V_2$.
The divergence degree is at most zero. The significant case
occurs when $\omega(\cG)=0$ and for that one must have 
\beq  C_{\bG} = 1, \quad g_{\bG} =  \sum_{\bJ} g_{\bJ} =0,  
\quad V_2 = 0.
\eeq

\noindent{\bf Case} $N_{\ext} = 2$:  
Since $ C_{\bG} \leq N_{\ext}/2$, we should
have $ C_{\bG}=1$ and, therefore, $P(\cG)=  V_2 -1$. 
Furthermore, the only possible boundary colored graph
made with two vertices is unique and does not have
any genus $g_{\bG}=0$. We conclude that
the divergence degree is at most 1.
One may have $\omega(\cG)=1$ if $P(\cG)=-1$, and only if 
\beq  C_{\bG} = 1, \quad g_{\bG} =\sum_{J} g_{\tJ} = 0,  \quad V_2 = 0.
\eeq
Besides, $\omega(\cG)=0$ can happen in two cases: (1) $P(\cG)=0$, in which case, we have
\beq  C_{\bG} = 1, \quad g_{\bG} =\sum_{J} g_{\tJ} = 0,  \quad  V_2 = 1
\eeq
or (2) $P(\cG)=-1$ and, so,
\beq  C_{\bG} = 1\;, \quad g_{\bG} =0, \quad \sum_{J} g_{\tJ} = 1, 
\quad V_2= 0\;.
\eeq
In summary, the divergent graphs are determined by the Table 1
of Subsection \ref{subsect:digraph}.

\section{Formal sum approximations}
\label{app:formal}

We start by proving the claim \eqref{sum2} by giving the relevant part of the formal sum
\bea
S &=&
\sum_{p_1, p_2 \in \N } \frac{1}{( p_2 + p_3 +A )^2} 
 = \sum_{p=0}^\infty \frac{(p+1)} {(p + A)^2 }, \crcr
S_i &=& 
\sum_{p = M^{i}}^{M^{i+1}}
 \frac{(p+1)} {(p + A)^2 }  
= 
\frac{1}{A+M^i} 
+ \psi^{(0)}\left(M^{i+1}+A+1 \right)-\psi^{(0)}\left(M^i+A+1\right) \crcr
&&-
   (A-1) \left(\psi^{(1)}\left(M^i+A\right)-\psi^{(1)}\left(M^{i+1}+A
+1\right)\right)
\label{sum11}
\eea 
where $A$  is a constant (which is set $\frac{m}{a_2}$
in the text) and $\psi^{(n)}(z)$ is the so-called polygamma function 
or $n^{th}$ derivative of $\ln \Gamma[z]$. 
There are well-known relations satisfied by 
the polygamma functions and are given by
\bea
&&
\psi^{(m)}(1+ z)  =\psi^{(m)}( z)   + (-1)^{m} m!\, z^{-(m+1)} ,\crcr
&&
\psi^{(0)}(z) \sim  \ln z - \frac1{2z}  + \dots ,    \qquad z \to \infty , \qquad 
 |\arg z| < \pi, \crcr
&&
\psi^{(n)}(z) \sim (-1)^{n-1} \left[ \frac{(n-1)!}{z^n} +\frac{n!}{2 z^{n+1}}
+ \dots \right] ,   \;\; z \to \infty , \;\;
 |\arg z| < \pi, \;\; n \geq 1, 
\eea
so that the divergent part of \eqref{sum11} as 
$M^i \gg 1$  can be given as (neglecting all factors of the form 
constant of convergent factors of the form $1/M^i$ or
constants)
\bea
S_i \sim 
 \kappa  [(i+1) \ln M- i\ln M] = \kappa \ln M
\label{sumfin}
\eea  
hence this does not depend on the scale $i$.

\section*{Acknowledgements}
D.O.S. thanks the Perimeter Institute for its hospitality. 
Discussions with V. Bonzom, R. Gurau and V. Rivasseau 
are gratefully acknowledged. 
Research at Perimeter Institute is supported by the Government of Canada through Industry
Canada and by the Province of Ontario through the Ministry of Research and Innovation.

\end{document}